\documentclass[twocolumn,showpacs,preprintnumbers,amsmath,amssymb,rmp]{revtex4}

\usepackage{latexsym}
\usepackage{graphicx}

\begin{document}
\title{How to find decision makers in neural circuits?}
\author{Alexei A. Koulakov$^{1}$, Dmitry Rinberg$^{2}$, and Dmitry N. Tsigankov$^{1}$} 

\address{
\protect{$^1$}\hspace{-.0in}Cold Spring Harbor Laboratory, Cold Spring Harbor, NY, USA \\
\protect{$^2$}Monell Chemical Senses Center, Philadelphia, PA, USA
}

\begin{abstract}
Neural circuits often face the problem of classifying stimuli into discrete 
groups and making decisions based on such classifications. Neurons of these 
circuits can be distinguished according to their correlations with different 
features of stimulus or response, which allows defining sensory or motor 
neuronal types. In this study we define the third class of neurons, which is 
responsible for making decision. We suggest two descriptions for contribution of 
units to decision making: first, as a spatial derivative of correlations between 
neural activity and the decision; second, as an impact of variability in a given 
neuron on the response. These two definitions are shown to be equivalent, when 
they can be compared. We also suggest an experimental strategy for determining 
contributions to decision making, which uses electric stimulation with time-
varying random current. 
\end{abstract}

\pacs{n/a}

\maketitle

\section{INTRODUCTION}\label{introduction}

Nervous system is continuously confronted by megabytes of information, 
representing light, sound, smell, etc. This information is compiled by the 
brain into a set of decisions, representing behaviors of living organisms. 
The mechanisms involved in this reduction have been under investigation for 
many years (Glimcher, 2003; Romo and Salinas, 
2003). In this study we address a question complimentary to the issue of 
decision making (DM) mechanisms. We define neuronal units involved in making 
perceptual decisions. For this purpose we determine DM activity in surrogate 
networks, defined mathematically, in which a complete control is present 
over stimuli, mechanisms, and responses. Such decision making analysis (DMA) 
has practical significance, since once units involved in making particular 
decision are located, further efforts could be concentrated on uncovering 
the underlying mechanisms. 

In this study DM task is defined as evaluation of a function in the 
multidimensional stimulus space (Figure 1A). This function has a discrete set 
of values, representing the repertoire of responses available to the 
organism. The decisions may, of course, be stochastic, to reflect the 
uncertainty, pertinent to behavior. This definition is suitable for 
experiments where subjects perform poly-alternative forced-choice tasks, 
such as saccadic response to the direction of stimulus motion 
(Shadlen and Newsome, 2001).

\begin{figure}[htbp]
\centerline{\includegraphics[width=3.0in]{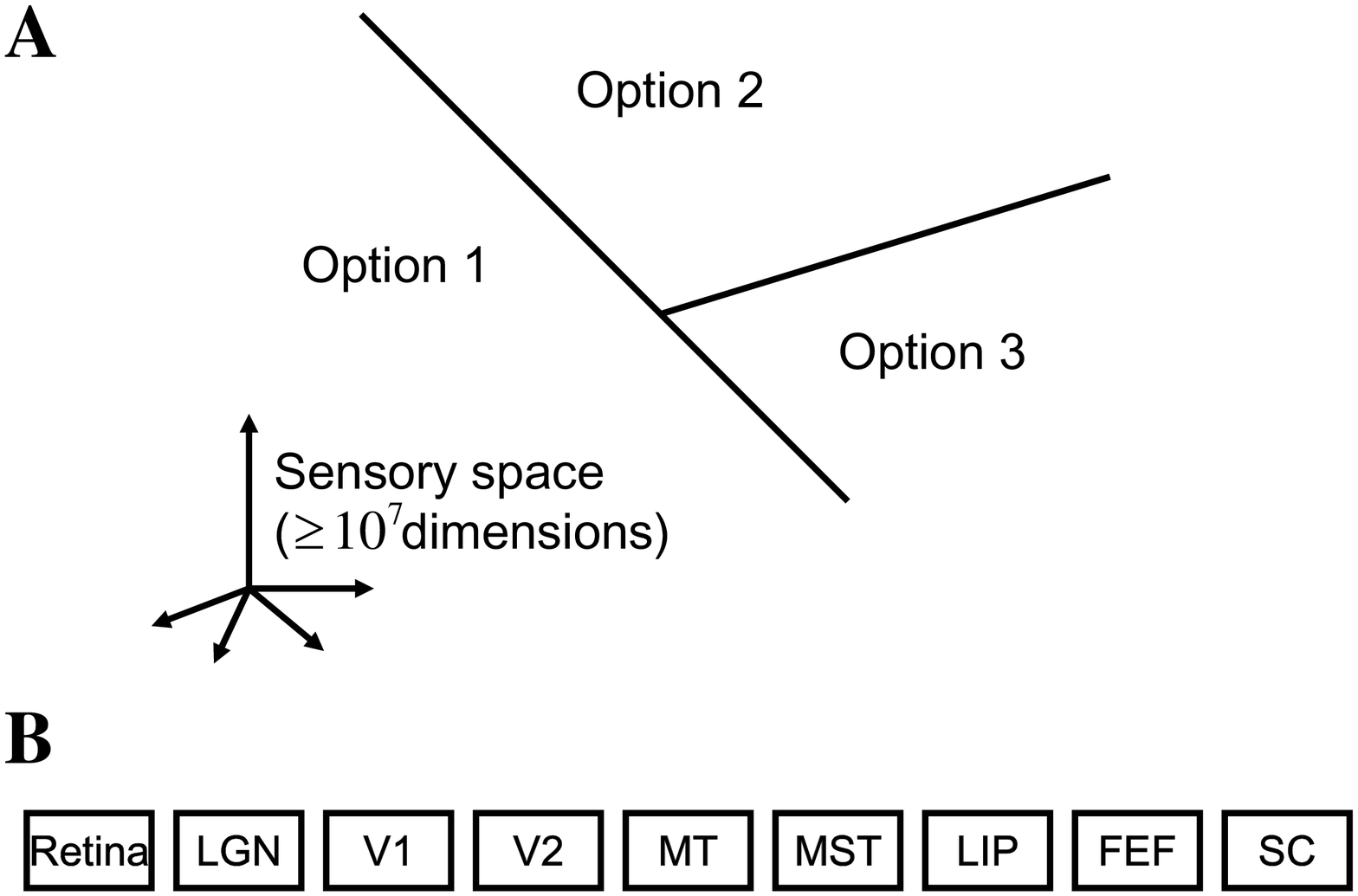}}
\caption{\textbf{A,} Definition of decision making task. Nervous system 
evaluates a function, whose values represent discrete decisions, in the 
many-dimensional sensory space. \textbf{B,} Some of the visual areas 
involved in motion-discrimination task. The areas on the left are more 
sensory (response is correlated with the sensory input), while those of the 
right are more motor (correlated with the response).
}
\label{fig1}
\end{figure}

Let us consider motion-discrimination task in more detail. Figure 1B lists 
some visual areas, which are involved in this task. The areas are arranged 
along a rough sensory-motor axis, so that the areas on the left are more 
``sensory'', while those on the right are more ``motor''. This implies that 
the responses in these areas are more correlated with stimulus or response 
respectively. Where on this sensory-motor axis one should position the DM 
elements? One could argue that the elements most correlated with the 
decision itself are the decision makers, following the analogy with the 
definition of sensory and motor elements. It is, however, difficult, if not 
impossible, to distinguish such definition from the definition of purely 
motor units (Shadlen and Newsome, 2001). The latter 
relay the results of decision making process, without involvement in the 
formation of the decision. An alternative approach is therefore needed to 
define the DM units. 

The DM components may be surmised to be located on the interface between 
sensory and motor areas. More precisely, the \textit{first} element in the sensory-motor 
chain, which carries significant correlation with the response, may be 
identified as the decision maker. In this study we develop this idea into 
rigorous mathematical formulation and find a special correlation function, 
which determines contributions of units to DM. This 
formalism allows us to answer two questions pertaining to the identities of 
DM units. First, we consider the case when not one but \textit{several} elements are 
involved in the same decision simultaneously. Our approach allows us to 
evaluate relative importance of various units in such a distributed DM. 
Second, we consider the systems with loops in connectivity. For such systems 
the concept of `the first element' becomes more arbitrary and one has to 
proceed more carefully in defining contributions to DM. We succeed in doing 
so for our surrogate networks and define DM units for recurrent networks in 
a way, which is consistent with the linear sensory-motor chains, thus 
satisfying the requirement of the correspondence principle. 

This paper is organized as follows. We first analyze simple linear chain models,
and networks, such as trees, which have similar properties. We then 
use this analysis to define decision makers in networks of arbitrary 
connectivity. Finally, we extend our study to the cases, when electric 
stimulation can be applied to units, and show that DM components can be 
identified in a way consistent with our preceeding analyses. 

\section{LINEAR CHAINS AND THEIR DERIVATIVES}
\label{linearchain}

The goal of this section is to formulate quantitative principles by which DM network elements can be identified. 
We approach this task by analyzing simple cases, which can be solved exactly without the use of computer, and in which the identities of DM elements are clear. 
These cases allow us to emphasize the properties of DM task we are attempting to describe.
We proceed therefore to the analysis of the simplest network capable of making decisions.

\subsection{The `nematode' network}
\label{nematode}

In this subsection we consider the network, which we call `nematode', because of its resemblance to simple biological organisms, both in the layout and in the fundamental significance.
We first define the model; then show that it can make simple decisions; and, finally, define the positions of decision makers in the network.

\begin{figure}[htbp]
\centerline{\includegraphics[width=3.0in]{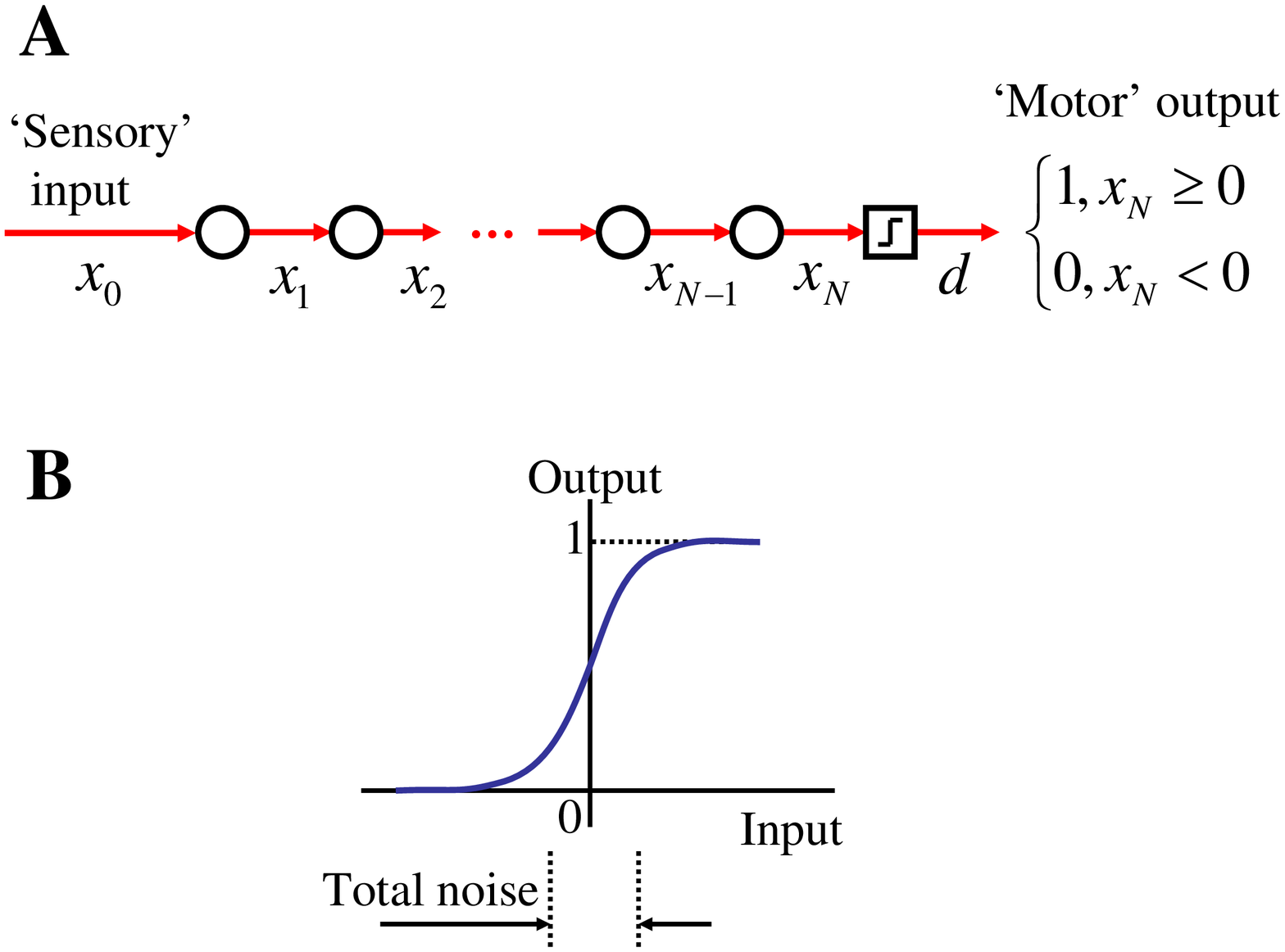}}
\caption{\textbf{A,} a simple `nematode' network consists of a linear chain 
of units. All units in the chain, but 
the last, are linear. The last unit, shown by the square, is non-linear and 
returns zero or one depending on the sign of the response of the preceding 
unit. \textbf{B,} The average input-output relationship for the `nematode' 
is given by the sigmoid function (error function). The spread of the sigmoid 
is determined by the net noise in the chain.
}
\label{fig2}
\end{figure}

Consider a linear chain of units, whose response is characterized by a set 
of real numbers $x_i$, where $i=1...N$ is the position of the element in 
the chain (Figure 2). Response of each element does not depend on time. This 
model is therefore static. This assumption is introduced here to simplify 
the analysis and can be relaxed as described below (section~\ref{trees}). Each unit 
performs a simple linear transformation between the unit's input and the 
output. Thus, for element number $i$
\begin{equation}
\label{eq1}
x_i =x_{i-1} +\eta _i 
\end{equation}
Here $\eta _i $ is noise associated with the element. In this work we assume 
that noise has zero mean, is individual to each unit, and, therefore, is 
uncorrelated between units, i.e.
\begin{equation}
\label{eq2}
{\begin{array}{*{20}c}
 {\overline {\eta _i } =0,} \hfill & \hfill \\
\end{array} }\overline {\eta _i \eta _j } =\left\{ {{\begin{array}{*{20}c}
 {\overline {\eta _i^2 } ,} \hfill & {i=j} \hfill \\
 {0,} \hfill & {i\ne j} \hfill \\
\end{array} }} \right.
\end{equation}
We further assume that noise has a Gaussian distribution. The chain of 
linear elements is thus completely specified by a set of noise variances 
$\overline {\eta _i^2 } $. The model described by (\ref{eq1}) and (\ref{eq2}) yields the 
following solution for the response of the last element in the chain 
\begin{equation}
\label{eq3}
x_N =x_0 +\eta _1 +\eta _2 +...+\eta _{N-1} +\eta _N .
\end{equation}
Thus, the response of the last element is just a sum of the input into 
network $x_0$ and noise contributions from all units, independently on the 
order of unit in the chain. 

The last element in the chain has non-linear response properties. Its 
response is defined by
\begin{equation}
\label{eq4}
d=H(x_N ),
\end{equation}
where $H(x)$ is the Heaviside~step~function, which is equal to one/zero if 
the argument is positive/negative. It follows then that our `nematode' 
network is capable of making decisions based on the values of input 
variable $x_0$. This is if we interpret variable $d$, which is equal either 0 
or 1, as the result of DM process, as defined in Figure 1A. The decisions 
are made stochastically and are dependent upon the instantiations of random 
variables $\eta _i $, which vary from trial to trial. 

\begin{figure}[htbp]
\centerline{\includegraphics[width=3.0in]{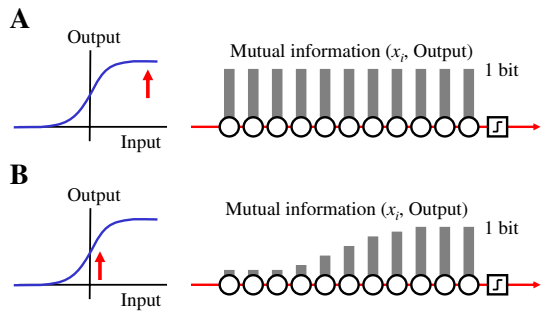}}
\caption{
\textbf{A,} If signal-to-noise ratio is high, responses of all 
units are well correlated with the output, as shown on the right by the 
mutual information between the response of given unit and the output. 
\textbf{B,} For the case of low signal-to-noise ratio, the output is more 
correlated with the motor units (right) than with the sensory ones (left).
}
\label{fig3}
\end{figure}

Our model is completely defined by the set of noise variances, pertinent to 
each unit $\overline {\eta _i^2 }$. Although decisions made by this chain 
are quite simple, the identities of decision makers are not so easy to find. 
The distribution of impact to DM along the chain should depend upon the 
distribution of noise variables $\overline {\eta _i^2 } $. Our next goal is 
to develop a sensible definition of contributions to DM based on the vector 
of variances $\overline {\eta _i^2 } $. Before doing so we describe general 
input-output properties of the chain. 

Since decision made by the network varies from trial to trial, one can 
define averaged over trials response of the system $\overline {d(x_0 )} $. 
As shown in Figure 2B it has a sigmoid shape, smeared by the total amount of 
noise in the system. One can, therefore, consider two cases, depending on 
whether the signal-to-noise ratio for the chain is large or small. These two 
regimes are shown in Figure 3A and B respectively. 

To analyze responses of units in these two cases we define their correlation 
with the decision. This correlation is defined for each element in the chain 
(Figure 3, right). As a measure of correlation we choose mutual information 
(MI) between response of the $i-$th unit, $x_{i}$, and the decision, $d$. MI has an 
advantage of being unitless (it is measured in bits) and having clear 
intuitive properties, as described below. We will also show below in this 
section that MI has limitations as a measure of DM.

MI describes the information transmission from the $i-$th unit to the output of 
the system. Since the output can only have values 0 or 1, MI cannot exceed 
the value of one bit. We now consider two cases, depending on the network's 
signal-to-noise ratio. If network input $\left| {x_0 } \right|$ is large, as 
in Figure 3A, response of the system is well correlated with the input. 
Hence, activities of all units are well correlated with both input and 
output, and $MI(x_i ,d)\approx 1$ for all of the units. In the opposite 
limit, when the signal-to-noise ratio is small, $\left| {x_0 } \right|$ is 
smaller than noise, and the system's response is weakly correlated with the 
input (Figure 3B). In this case MI as a function of unit's position displays a 
structure, shown in Figure 3B (right). This structure, as shown below, has a 
key to the definition of DM components and is qualitatively discussed here. 
The units, which are close to the exit from the network, show strong 
correlation with the decision, similarly to the high signal-to-noise ratio case. 
Their MI is therefore close to 1 bit. On the 
other hand, more `sensory' units, in the beginning of the chain are strongly 
correlated with the input. Since input-output correlation is weak in low 
signal-to-noise ratio case, the `sensory' units display virtually \textit{no} relation 
to the output and $MI(x_i ,d)\approx 0$ for such units (Figure 3B, right). 
Thus, MI, as a function of $i$ displays a transition from 0 to 1 in the
low signal-to-noise ratio case. 

How could one deduce identities of decision makers from these dependencies 
(Figure 3A and B)? One could suggest that the elements perfectly correlated 
with the output of the system, such as exit elements from the chain, are the 
ones that make the decision. However, such elements may be just the relay or 
`motor' units, in which case their contribution to DM is small. Indeed, when 
we type, our decisions are perfectly correlated with activities of finger 
muscles; but one could hardly blame our fingers for the content of the 
typing. Thus, despite their high correlation with the output, exit elements 
could not be called decision makers. Input elements, having no correlation 
with the decision, are responsible for DM in even lesser degree. We thus 
need to analyze the dependence of MI on position in more detail and suggest 
another scheme for defining DM units.

Our discarding of motor units as decision makers can be further extended 
onto the entire high signal-to-noise ratio case (Figure 3A). We suggest that 
the deterministic regime is not descriptive from the point of view of DM 
analysis. First, in this regime all units become indistinguishable from 
motor. The latter are not decision makers, as suggested above. Second, the 
dependence shown in Figure 3A (right) does not reveal the contributions of 
individual units to the decision. Since all units have the same correlation, 
it is hard, if not impossible, to differentiate them and assign different 
contributions. Third, the responses of units in this case are 
deterministically related to the input. Hence, units act as relays, 
passively transmitting information along the chain. It can be argued that 
the external environment, providing the input variable $x_0$, acts as the 
decision maker. We conclude that to find decision making activity one has to 
concentrate on the low signal-to-noise ratio case. 

We show below that the identities of decision making units can be deduced 
from the shape of transition in Figure 3B (right). To this end we analyze a 
set of examples of networks with various distributions of noise $\overline 
{\eta _i^2 } $. We start from the simplest example of a single noisy unit.

\subsubsection{Example 1: 'Noisy' neuron.} 

Consider a chain in which noise is absent from all units but one, whose 
order number in the chain is $n$ (Figure 4). Since, according to our 
previous discussion, we need to consider the low signal-to-noise ratio case, 
we will assume that
\begin{equation}
\label{eq5}
x_0 =0,
\end{equation}
i.e. network receives no input. Making the decision in this case is still possible, 
based on the values of noise inside the network. Since noise is only present in one neuron, from 
(\ref{eq3}) we conclude that 
\begin{equation}
\label{eq6}
x_N =\eta _n .
\end{equation}
The decision made by the network is
\begin{equation}
\label{eq7}
d=H(\eta _n ).
\end{equation}
Thus, decision is causally linked to the processes controlling unit number 
$n$, which leads us to conclusion that this neuron is the decision maker. 

Paradoxically, the noisiest unit in this simple formulation makes the 
largest impact. All noiseless elements, even nonlinear, are deterministic, 
and work as simple relays which transmit information from the previous node 
to the next one. The output of the circuit is linked to the processes 
controlling noise in neuron number $n$, rather that in any other neuron in 
the network.

One would be tempted to conclude that the non-linear element is actually the 
decision maker in this case. We deduce that the non-linear element does 
not have a causal effect on output from the circuit; therefore its role is 
just to relay response from neuron $n$ to the output. In this respect the 
non-linear element is not different from other noiseless elements. 

To link this example to our previous discussion (Figure 3B) we plot MI as a 
function of position in the chain in Figure 4 (top). As we discussed, MI is 
high for exit (`motor') units and low for input (`sensory') elements. Figure 
4 also shows the derivative of MI with respect to position in the chain. It 
is clear that this derivative represents the decision making element. Thus, 
we conclude that not correlation with the decision but the \textit{rate of change} of the latter 
along the network is the indicator of DM.

\begin{figure}[htbp]
\centerline{\includegraphics[width=3.0in]{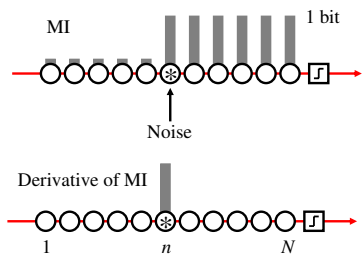}}
\caption{
The example of 'noisy' neuron (marked by asterisk). Top 
panel, mutual information between given unit and the decision. Bottom panel, 
derivative of mutual information. The derivative represents the 
decision making unit in this case. 
}
\label{fig4}
\end{figure}

\subsubsection{Example 2: Uniformly distributed noise.} Our next example shows that the conclusion about derivative of MI is 
basically correct, but has to be slightly amended to be numerically precise. 
Consider the chain in which all elements are noisy and the variance of noise 
is the same for each element. In this case 
\begin{equation}
\label{eq8}
x_N =\eta _1 +\eta _2 +...+\eta _{N-1} +\eta _N 
\end{equation}
i.e. all units contribute to decision \textit{equally}. This is because Eq. (\ref{eq8}) does not 
distinguish the order in which contributions from the units are added, and 
all contributions are of equal strength on average. Can this conclusion be 
confirmed by the derivative of MI?

Figure 5A shows MI as a function of position in the chain for this case. 
This dependence is obtained in Appendix A. It increases smoothly from 0 to 1 
resulting in a non-zero derivative at all units. This is consistent with 
(\ref{eq8}) and the notion that all units participate in the decision. However, 
(\ref{eq8}) suggests that all units participate in decision \textit{equally}. The derivative of MI 
turns out to be slightly non-uniform, as seen in Figure 5A. This can be 
corrected if not MI itself but a non-linear function of MI, denoted $F(MI)$, 
is considered. This non-linear function is calculated in Appendix A and is 
shown in Figure 5B. The new correlator $F(MI)$ has the same basic properties 
as the MI. It rises from 0 to 1 monotonously when passing through the array 
(Figure 5C). But, in addition, its derivative turns out to be \textit{uniform}, as shown in 
Figure 5C (bottom). This is consistent with equal participation of all units 
in DM in the uniformly distributed noise case and Eq.~(\ref{eq8}). Thus, we 
conclude that for this case the contributions to DM are given by the rate of 
increase of $F(MI)$ when moving through the array
\begin{equation}
\label{eq9}
DM_i =F\left( {MI_i } \right)-F\left( {MI_{i-1} } \right).
\end{equation}
Here $i$ is the index along the chain. Eq.~(\ref{eq9}) is the main result of this 
paper. It represents our definition of contributions to DM for networks of 
simple connectivity, such as chains. 

Three points should be made about the definition (\ref{eq9}). First, it reproduces 
the result obtained in the previous example of 'noisy' neuron. Indeed, the 
mutual information rises from 0 to 1 on the 'noisy' neuron in Figure 4. But 
$F(MI)$ coincides with MI at these values, as follows from its plot in 
Figure 5B. Thus, the derivative of $F(MI)$ is also given by a single spike 
at the position of 'noisy' neuron, as in Figure 4 (bottom). Second, Eq. (\ref{eq9}) 
implies that, from point of view of DM, not mutual information, but another 
correlator, given by $F(MI)$, is more relevant. Function $F$ deviates from 
linear function only slightly (Figure 5B), and for practical purposes the 
distinction between the MI and $F(MI)$ could be ignored. However, we retain 
it throughout the manuscript to ensure mathematical rigor. Third, when 
deducing (\ref{eq9}) we did not postulate that contributions to DM are 
proportional to the variance of noise. Instead, we suggested that Eq. (\ref{eq8}) 
implies that all units contribute equally, independently on the order in the 
chain. This simple qualitative statement is powerful enough to constrain our 
quantitative reasoning and lead to a measure of DM in form of function 
$F(MI)$ and definition (\ref{eq9}). We do not know yet if the derivative of 
$F(MI)$ is proportional to the variance of noise, square root of this 
variance, or any other characteristic of noise in each element. All of these 
parameters give the same results in the uniform noise case. We need to have 
a difference between units to measure relative strength of their 
contributions. This is achieved by the next example.

\begin{figure}[htbp]
\centerline{\includegraphics[width=3.0in]{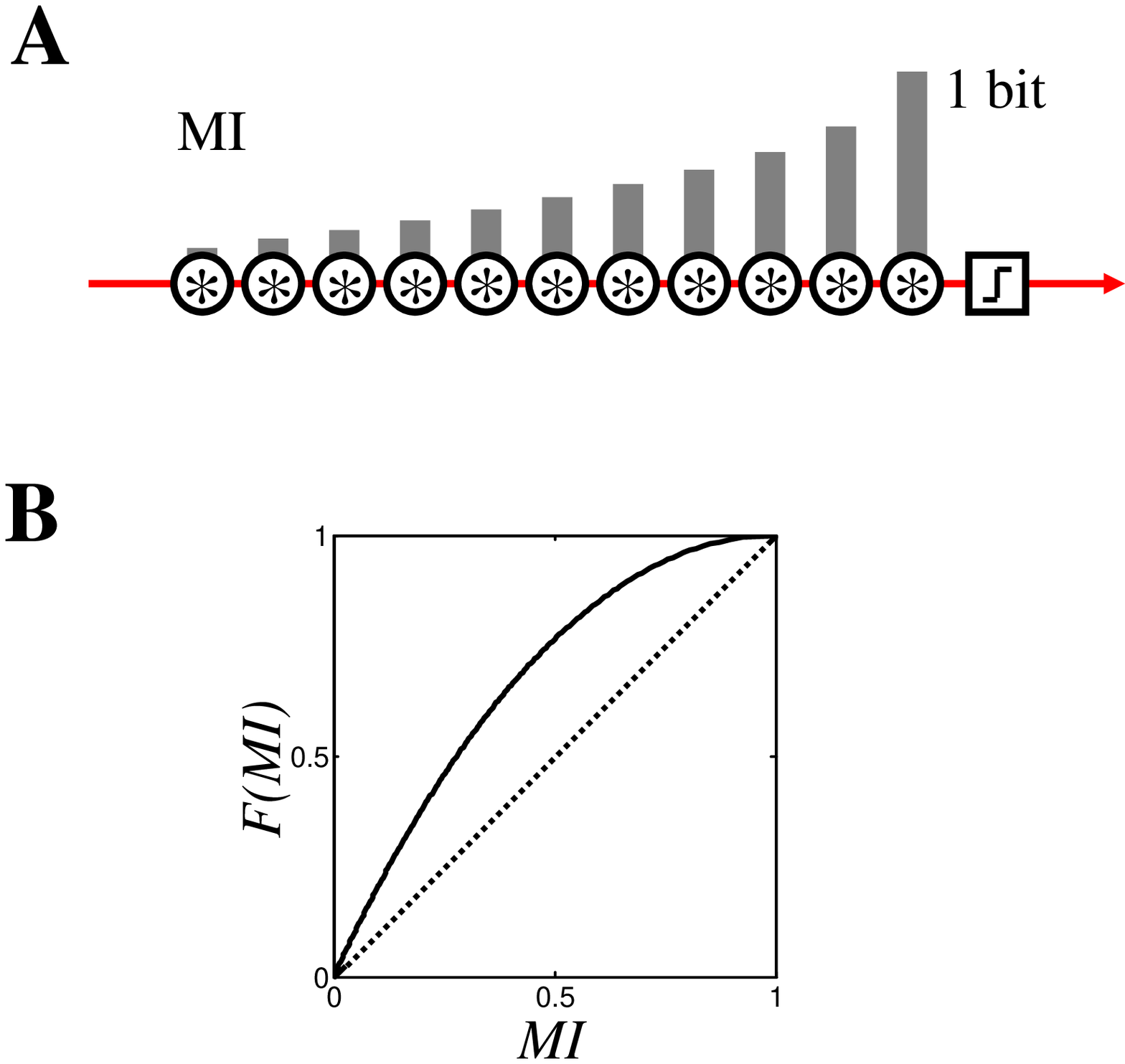}}
\centerline{\includegraphics[width=3.0in]{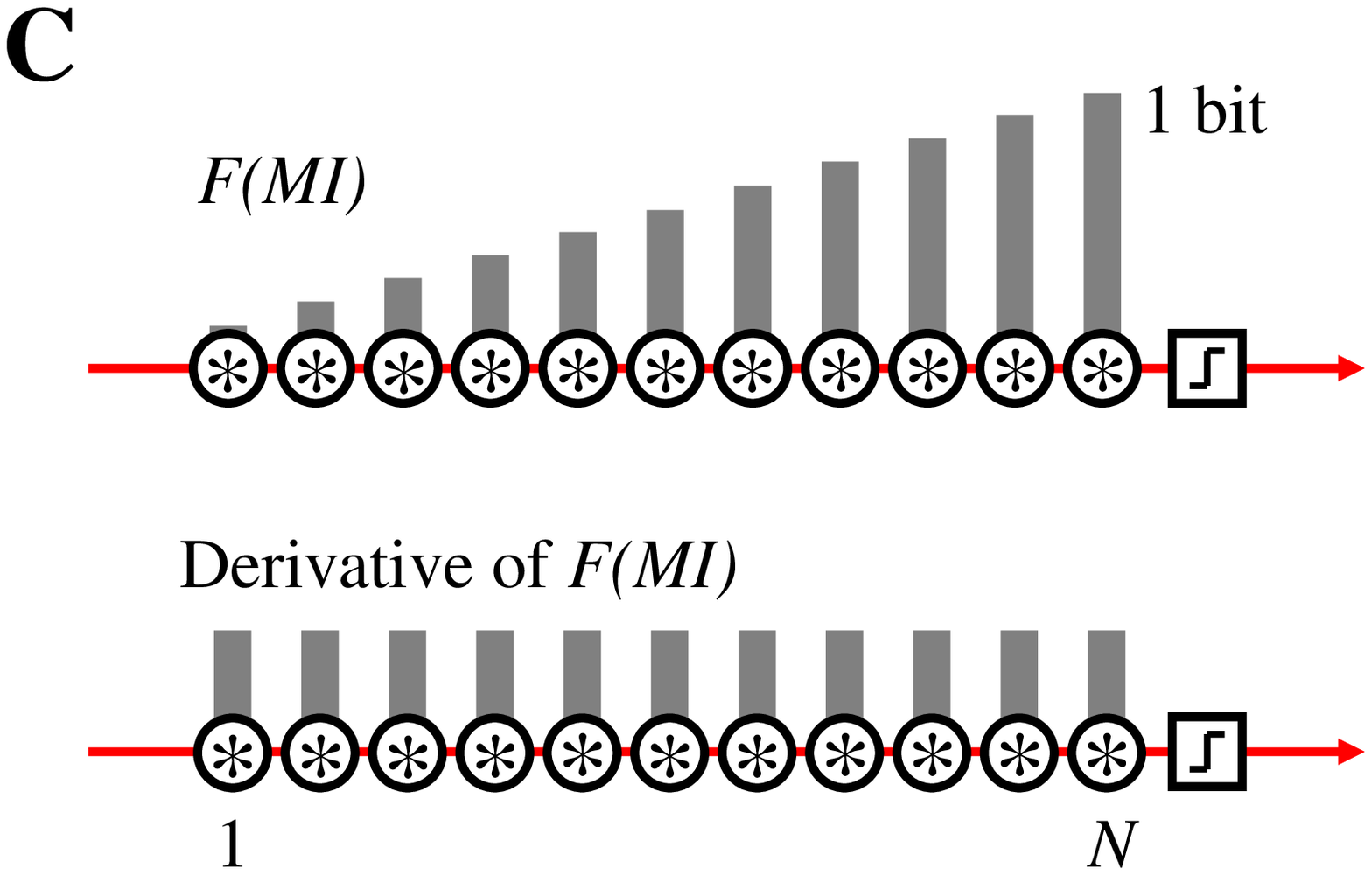}}
\caption{
The example with uniformly distributed noise. 
\textbf{A}, mutual information between response of given unit and the 
decision. The dependence has a non-uniform increase, suggesting that mutual 
information is not a good measure of decision making. \textbf{B}, if one 
applies a non-linear function (solid curve) to the mutual information in 
\textbf{A}, one obtains a uniformly increasing correlator in \textbf{C}. 
This non-linear function, called $F(MI)$, is close to linear, shown by the 
dotted line. \textbf{C}, the new correlator $F(MI)$ (top panel) has a 
uniform derivative (bottom panel). Thus, derivative of $F(MI)$ is a sensible 
measure of decision making in the case of uniform noise. 
}
\label{fig5}
\end{figure}

\subsubsection{Example 3: `Loud' neuron.} In this example the variances of noise on all neurons are the same, 
similarly to the previous case. However, here we amend the network 
definition given by (\ref{eq1}). We do so for only one neuron. We assume that the 
link between units 5 and 6 is characterized by a very large strength $K>>1$. 
Thus, for neuron number 6 (Figure 6) instead of (\ref{eq1}) we have
\begin{equation}
\label{eq10}
x_6 =Kx_5 +\eta _6 
\end{equation}
Therefore this example is the same as the previous, except that the single 
network connection is changed. What are the DM units in this case?

The network's output is given by 
\begin{equation}
\label{eq11}
x_N =K(\eta _1 +\eta _2 +...+\eta _5 )+\eta _6 +...+\eta _{11} 
\end{equation}
Thus, units 1 through 5 contribute equally to decision. In addition, their 
contributions are multiplied by a large factor $K$. Units 6 through 11 also 
contribute equally, but their contribution is much smaller than that of the 
former group. We conclude that units 1 through 5 are much stronger 
decision makers than units 6 through 11. This conclusion is supported by the 
derivative of $F(MI)$, as shown in Figure 6 (bottom).

\begin{figure}[htbp]
\centerline{\includegraphics[width=3.0in]{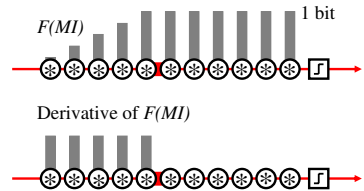}}
\caption{
The `loud' neuron example. The link between units 5 an 6 
is strengthened. Compare to Figure 5A.
}
\label{fig6}
\end{figure}

Thus, changing one link in the chain produces large effect on the 
distribution of DM. The units downstream from the link contribute less to 
decisions, while the units upstream contribute a lot. What is the measure of 
decision making, which could differentiate these two types of units? 

Calculations in Appendix A show that derivative of $F(MI)$ is proportional 
to $K^2$ for units 1 through 5. This is easy to understand qualitatively, 
since MI increases along the chain even for negative ($K<0)$ links. This 
is \textit{not} possible if contribution from units 1 to 5 are multiplied by $K$ for 
example. Thus, an even power of $K$ is required, which is shown in Appendix 
A to be $K^2$. 

\subsubsection{Alternative definition of DM.} So far we have used definition (\ref{eq9}), which is quite complex, since it 
involves calculation of a nonlinear function $F(MI)$. Is it possible to 
reproduce the results derived above in a simpler way? It turns out that the 
role of given unit in DM is proportional to its contribution to the 
variability of the output $\overline {x_N^2 } $. This leads us to an 
alternative to (\ref{eq9}) definition of DM. 

Let us introduce the new definition using the examples, considered above. 
From (\ref{eq11}) in the `loud' neuron case we derive 
\begin{equation}
\label{eq12}
\overline {x_N^2 } =K^2(\overline {\eta _1^2 } +...+\overline {\eta _5^2 } 
)+\overline {\eta _6^2 } +...+\overline {\eta _{11}^2 } .
\end{equation}
We could conjecture that the contributions to DM from different units are 
weighted proportionally to the corresponding summands in (\ref{eq12}). Indeed, if 
we assume
\[
DM_{1..5} =\overline {\eta _{1..5}^2 } K^2
\]
\begin{equation}
\label{eq13}
DM_{6..11} =\overline {\eta _{6..11}^2 } ,
\end{equation}
by choosing appropriate values of variance of noise and gain, we can 
reproduce the results of all three of our previous examples. Thus, in the 
case of 'noisy' neuron the variance of noise is only present in one unit, 
rendering this unit decision maker, according to (\ref{eq13}). In the case of 
uniform noise, when $K=1$ and all $\overline {\eta _{1..11}^2 } $ are the 
same, (\ref{eq13}) gives uniform contributions to DM. In the case of `loud' neuron, 
(\ref{eq13}) gives the correct factor $K^2$ describing the advantage of upstream 
neurons. Thus, the contributions to DM are proportional to the variance of 
noise on given element, multiplied by the square of the gain from this 
element to the output. We can rewrite (\ref{eq13}) in a more compact form to 
emphasize this latter statement
\begin{equation}
\label{eq14}
DM_i =\overline {\eta _i^2 } \frac{d\overline {x_N^2 } }{d\overline {\eta 
_i^2 } }.
\end{equation}
One could verify (\ref{eq14}), by applying it to (\ref{eq12}) and obtaining 
relationships (\ref{eq13}). This justifies (\ref{eq14}) in the three examples considered 
above.

Eq. (\ref{eq14}) also applies to linear chains in general. In Appendix A we derive 
(\ref{eq14}) from previous definition (\ref{eq9}) for arbitrary distribution of 
connection strengths and noise. Thus, (\ref{eq14}) can be considered an 
alternative definition to (\ref{eq9}). The equivalence between (\ref{eq9}) and (\ref{eq14}) is 
demonstrated graphically in Figure 7. 

Why should one consider an alternative definition? This is because (\ref{eq9}) 
cannot be applied to networks of arbitrary connectivity, such as circuits 
containing loops. Definition (\ref{eq14}) however applies to all topologies, 
including the linear chain examples, considered here. 

\begin{figure}[htbp]
\centerline{\includegraphics[width=3.0in]{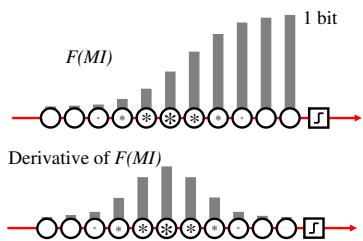}}
\caption{
Equivalence of two definitions. The top panel shows 
distribution of noise variance (asterisk diameter) and of $F(MI)$ (bars). 
The bottom panel displays the derivative of $F(MI)$, defined by (\ref{eq9}). The 
derivative is numerically the same as the variance of noise. Both can be 
used as measures of decision making.
}
\label{fig7}
\end{figure}

\subsection{Conclusions from `nematode' study}
\label{conclusions}

Let us review our findings. First, we arrived to the definition of DM 
activity using the information-theoretical approach (\ref{eq9}). According to this 
definition, DM is the rate of change of correlation with decision along the 
chain. In other words, the \textit{first} element or elements, which correlate with the 
decision, are the decision makers. This approach has its pros and contras. 
Indeed, the viewpoint expressed by (\ref{eq9}) has a potential to be transferred 
to other systems, which contain non-linear elements. Eq. (\ref{eq9}) has an 
information-theoretical origin; hence its applicability may be broader than 
our simple system. 
%We emphasize, however, that this point is not confirmed 
%here, it will be further explored elsewhere. 
Another advantage of (\ref{eq9}) is 
that it relies on the characteristics measurable in single-electrode 
recording experiments, such as response of single unit and its correlation 
with behavioral decision. Thus, (\ref{eq9}) could be used experimentally. The 
disadvantage of the information-theoretical approach is that it is not clear 
how to apply it to the systems with loops, as we have mentioned above. Since 
biological networks almost always contain loops this significantly limits 
the applicability of information-theoretical formula (\ref{eq9}). 

Our second step was to derive an alternative definition (\ref{eq14}). The latter 
is \textit{equivalent} to the former definition (\ref{eq9}) for linear-chain (`nematode') example, as 
we have demonstrated on simple examples and have shown more rigorously in 
Appendix A. The alternative definition (\ref{eq14}) can be understood on the basis 
of the following two observations. First, the example of `noisy' neuron 
shows that the variability is the source of decisions. Thus,

\underline {\textbf{Conclusion 1:}} Under fixed other conditions, an 
increase in variability and noise in a single unit leads to a larger 
contribution to DM from this unit. 
\begin{equation}
\label{eq15}
DM_i \sim \overline {\eta _i^2 } 
\end{equation}

Second, the example of `loud' neuron shows that not only variability and 
noise are important but also how much of this variability reaches the motor 
units. DM is hence a property of network connectivity too. Thus, we arrive 
to the next rule

\underline {\textbf{Conclusion 2:}} The stronger is the pathway from given 
unit to the motor output, the larger is the contribution of this unit to DM. 
\begin{equation}
\label{eq16}
DM_i \sim \frac{d\overline {x_N^2 } }{d\overline {\eta _i^2 } }
\end{equation}
These two rules are combined into the definition (\ref{eq14}). Although (\ref{eq14}) and 
(\ref{eq16}) assume that the output element is unique, this requirement will be 
removed below, when we consider arbitrary topology networks.

What are the features of (\ref{eq14})? It could be used for an arbitrary topology 
network, since it does not contain derivative along the chain, as (\ref{eq9}) 
does. Definition (\ref{eq14}) can also be used operationally to measure the 
contribution of each neuron to the decision experimentally. To do that one 
needs to vary noise at the given unit and measure the variability of the 
responses. The details are discussed in section ~\ref{stimulation}
below.

A special note should be made about normalization in (\ref{eq14}). Throughout this 
work we adopt the convention that DM contributions are evaluated for all 
units and then normalized proportionally to (\ref{eq14}), so that the total sum of 
all contributions is equal to one (or 100{\%}). We will assume this to hold 
below without explicitly mentioning. Finally, we give another 
definition of DM contributions, which could be useful when noise in the 
system is the same for all units. In this case the only difference between 
units is due to difference in their position in the network. We therefore 
call such quantity \textit{topological} DM. 
\begin{equation}
\label{eq17}
TDM_i =\frac{\partial \sigma ^2(x_N )}{\partial \overline {\eta _i^2 } }
\end{equation}
As seen from e.g. (\ref{eq12}) it does not depend on the levels of noise, and can 
be obtained from (\ref{eq14}) by assuming that $\overline {\eta _i^2 } =1$ for all 
units. It therefore describes how strongly each elements of the circuit 
affects the output. This quantity is sometimes helpful in describing the 
network's topology.

Lastly, we discuss the notion of noise and variability in our approach. Is 
this really noise, which leads networks to decisions? Not necessarily. 
Imagine that we have studied a chain-like network (Figure 8A) and performed 
the DM analysis, described above. We found that the network contains two 
decision makers, which are equally important. A more thorough investigation 
may suggest that these units are inputs from external network, which in 
effect is responsible for DM. For example, these hidden pathways may be 
inputs from other sensory modalities or regulatory inputs of other type. 
Thus, DMA may help identify entry points from other, less studied, parts of 
the network.

\begin{figure}[htbp]
\centerline{\includegraphics[width=4.0in]{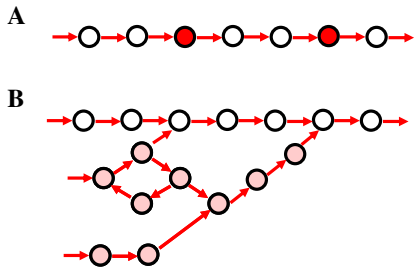}}
\caption{
Hidden pathway. The intensity of red shows 
contribution to decision making for each unit. \textbf{A,} analysis for an 
incomplete connectivity reveals two decision makers. \textbf{B,} a more 
thorough study may show that this results from other inputs to the network.
}
\label{fig8}
\end{figure}

\subsection{Trees}
\label{trees}

Our studies indicate that information-theoretical analysis [definition (\ref{eq9})] can be 
further extended to tree-like topologies (Figure 9). To this end we define 
column-vector 
$\mathord{\buildrel{\lower3pt\hbox{$\scriptscriptstyle\rightharpoonup$}}\over 
{f}}$ such that $f_i =F(MI_i )$. Then (\ref{eq9}) is equivalent to
\begin{equation}
\label{eq18}
\overrightarrow {DM} =(\hat {I}-\hat {S})\vec {f}.
\end{equation}
Here $\hat {S}$ is the structure matrix defined as follows. An element 
$S_{ij} $ of the structure matrix is equal to 1 if there is a connection 
from unit number $i$ to $j$. Matrix $\hat {I}-\hat {S}$ thus implements 
evaluating differences between connected elements in (\ref{eq9}). Structure matrix 
is related to connectivity matrix, containing network's weights through 
$S_{ij} =\left| {sign(C_{ij} )} \right|$. Connectivity matrices for some 
networks are shown in Figures 9 and 10.

\begin{figure}[htbp]
\centerline{\includegraphics[width=4.0in]{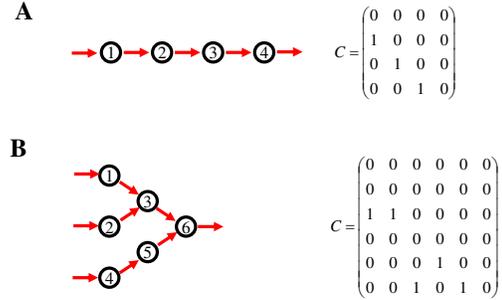}}
\caption{
Mutual information approach can be extended to 
connectivities other than linear chains (\textbf{A}). Thus, decision makers 
on trees (\textbf{B}) can also be found. Arbitrary network can be specified 
by connectivity matrices, which are provided for illustration purposes. The 
non-zero entries in a connectivity matrix indicate a connection between two 
elements numbered on the left. An entry value describes the strength of 
connection and does not have to be unitary or positive.
}
\label{fig9}
\end{figure}

Information-theoretical approach can be even further extended on the cases, when signals 
propagate along the network in time, therefore resulting in delays between 
signal and response. In this case by 
$\mathord{\buildrel{\lower3pt\hbox{$\scriptscriptstyle\rightharpoonup$}}\over 
{f}} $ one should understand a sum of correlations over all times preceding 
the decision. This compensates for the presence of delays. So far there is 
no understanding if Eq.~(\ref{eq18}) [or (\ref{eq9})] can be used for topologies other 
than trees. Definition (\ref{eq14}), however, can be used with networks of 
arbitrary connectivity. This is the topic of the next section.

\section{DYNAMIC MODELS}
\label{dynamic}

All previous examples, except the one mentioned at the end of the last 
session, were static, i.e. variables did not depend on time. The deficiency 
of this approach is that it is not clear how to treat networks with loops. 
To apply our analysis to the cases with loops, and, in general, to 
networks with \textit{arbitrary} connectivity (Figure 10), we consider time-dependent models 
here. This allows us to observe propagation of noise around the loop 
explicitly and to make accurate conclusions about contributions to DM. 

We limit ourselves to linear dynamical systems, where the single nonlinear 
element is the last one, transforming an analog system output to a binary 
response. As the first step we consider temporal dynamics in the 
discrete-time approximation, which contains all essential features of our 
approach. Later in the section we extend discrete model to the 
continuous-time case and show their equivalence.

\begin{figure}[htbp]
\centerline{\includegraphics[width=4.0in]{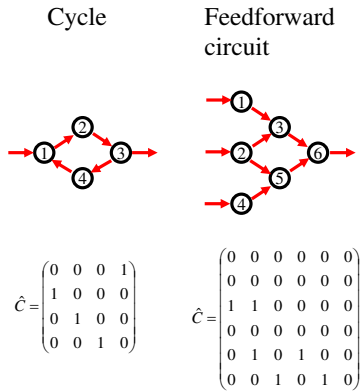}}
\caption{
The network topologies considered by the dynamic models. 
Arbitrary connectivities, such as cycles (left) or feedforward networks 
(right) can be considered. 
}
\label{fig10}
\end{figure}

\subsection{Discrete-time model}
\label{discrete}

In this section we consider a system of $N$ elements, whose activity at each 
instant is described by an $N$-dimensional column-vector $\vec {x}(t)$. Time 
has discrete values separated by an interval $\tau $. Therefore this model 
is called the discrete-time model. The values of activity at two neighboring 
time-slices are related by the connection matrix $\hat {C}$
\begin{equation}
\label{eq19}
\vec {x}(t+\tau )=\hat {C}\vec {x}(t)+\vec {\eta }(t)+\vec {s}(t)
\end{equation}
Here $\vec {\eta }(t)$ is the vector describing noise added to activity 
vector on each time-slice. The variable $\vec {s}(t)$ describes sensory 
input into the system. The rules of temporal evolution of activities 
described by this equation are general enough to include almost all 
interesting phenomena and mimic modeling of real systems on digital 
computers. In appendix B we will prove that this model is equivalent to systems with 
continuously defined time. 

Noise is specified by the parameter $\vec {\eta }(t)$, which has a zero mean 
and is defined by the correlation matrix 
\begin{equation}
\label{eq20}
\overline {\eta _i (t_1 )\eta _j (t_2 )} ={\cal N}_{ij} \delta _{t_1 ,t_2 } 
\end{equation}
We assume here that neighboring in time values on noise are not correlated, 
implying that we consider a system with white noise. This assumption can be 
easily relaxed and is used here to simplify the analysis. It becomes 
rigorously valid when time-interval $\tau $ is longer than the correlation 
time of noise. Further, if noise is specific to each neuron, the same-time 
correlation matrix $\hat {{\cal N}}$ is diagonal
\begin{equation}
\label{eq21}
{\cal N}_{ij} =\overline {\eta _i^2 } \delta _{ij} 
\end{equation}
This takes place i.e. when stochasticity is induced by probabilistic nature 
of synaptic vesicle release, in which case every two neurons receive 
uncorrelated fluctuating inputs. 

Some time after presentation of the stimulus [$\vec {s}(t)\ne 0$] the system 
is forced to make a decision through the following process. First, a scalar 
quantity 
\begin{equation}
\label{eq22}
y=\vec {v}^T\cdot \vec {x}(t)
\end{equation}
is evaluated. Here time corresponds to the instant, when the choice is to be 
made. The output metrics vector $\vec {v}$ describes the way in which 
system's activity affects motor response. In the simplest case, which was 
considered in the previous section, when a single element number $n$ evokes 
responses, $v_i =\delta _{in} $. In a more complex situation, when multiple 
areas/neurons have direct influence on decision, vector $\vec {v}$ has more 
than one non-zero element. On the second step, decision is made based on the 
sign of $y$
\begin{equation}
\label{eq23}
d=H(y)
\end{equation}
Thus, this model describes a two-alternative forced-choice task. 

Our system is completely defined by the following set of parameters: $\hat 
{C}$, $\vec {s}(t)$, $\hat {{\cal N}}$, and $\vec {v}$. As we have shown in 
the previous section, the presence of the stimulus is not required to define 
DM elements [Eq. (\ref{eq5})]. We therefore set $\vec {s}(t)$ to zero and are left 
with three parameters $\hat {C}$, $\hat {{\cal N}}$, and $\vec {v}$. We now 
are ready to determine DM elements in our simple model.

To find decision makers we will use Eq. (\ref{eq14}). In this case it becomes
\begin{equation}
\label{eq24}
DM_i ={\cal N}_{ii} \frac{\partial \sigma ^2(y)}{\partial {\cal N}_{ii} }
\end{equation}
Therefore, we need to evaluate the variability on the output from the system 
$\sigma ^2(y)$. This is accomplished if we notice that $y=\vec {v}^T\cdot 
\vec {x}(t)=\vec {x}^T(t)\cdot \vec {v}$ and 
\begin{equation}
\label{eq25}
\sigma ^2(y)=\vec {v}^T\cdot \overline {\vec {x}(t)\vec {x}^T(t)} \cdot \vec 
{v}=\vec {v}^T\hat {X}(t,t)\vec {v}
\end{equation}
Here we introduced the cross-correlation matrix defined as follows
\begin{equation}
\label{eq26}
\hat {X}(n,k)\equiv \overline {\vec {x}(n)\vec {x}^T(k)} =\overline {\left( 
{{\begin{array}{*{20}c}
 {x_1 } \hfill \\
 \vdots \hfill \\
 {x_N } \hfill \\
\end{array} }} \right)\left( {{\begin{array}{*{20}c}
 {x_1 } \hfill & \cdots \hfill & {x_N } \hfill \\
\end{array} }} \right)} 
\end{equation}
We replace here the time variable by the integers, specifying the time-slice 
number. The averaging in (\ref{eq25}) and (\ref{eq26}) is assumed over different 
instantiations of noise (trials). 

Due to the properties of noise in our model, this correlator does not depend 
on the absolute values of time ($n$ and $k)$, but only on the difference ($n - k)$. As 
follows from (\ref{eq25}), of particular interest is the same-time correlator $\hat 
{X}_0 \equiv \hat {X}(n,n)$, which determines fluctuations in $y$. We now 
derive equation for same-time correlator $\hat {X}_0 $.

Using (\ref{eq19}) we obtain
\begin{equation}
\label{eq27}
\begin{array}{l}
 \hat {X}_0 =\overline {\vec {x}(n+1)\vec {x}^T(n+1)} = \\ 
 =\overline {\left[ {\hat {C}\vec {x}(n)+\vec {\eta }(n)} \right]\left[ 
{\vec {x}^T(n)\hat {C}^T+\vec {\eta }^T(n)} \right]} \\ 
 \end{array}
\end{equation}
We then notice that the correlator $\overline {\vec {x}(n)\vec {\eta }^T(n)} 
$ is identically zero, since $\vec {x}(n)$ is a linear combination of values 
of noise at times $k<n$ [see Eq. (\ref{eq19})]. We thus deduce from Eq. (\ref{eq27}) that
\[
\hat {X}_0 =\overline {\hat {C}\vec {x}(n)\vec {x}^T(n)C^T} +\overline {\vec 
{\eta }(n)\vec {\eta }^T(n)} ,
\]
which leads us, finally, to 
\begin{equation}
\label{eq28}
\hat {X}_0 -\hat {C}\hat {X}_0 \hat {C}^T=\hat {{\cal N}}
\end{equation}
This equation allows us to determine the same-time correlator $\hat {X}_0 $ 
from connectivity and noise cross-correlogram, defined in (\ref{eq20}), which is a 
diagonal matrix.

We would like to pause here and describe the properties of this equation. 
First of all, in the most generic case (\ref{eq28}) allows us to determine $\hat 
{X}_0 $ from $\hat {C}$ and $\hat {{\cal N}}$ uniquely. Indeed, (\ref{eq28}) is a 
system of $N^2$ linear equations for $N^2$ unknowns $\hat {X}_0 $, arranged 
in the matrix form. Hence, this system, in most cases, can be solved 
uniquely. On the other hand, with one exception, $\hat {X}_0 $ cannot 
be expressed explicitly in terms of matrices $\hat {C}$ and $\hat {{\cal 
N}}$. Thus, one has to either appeal to the representation of $\hat {X}_0 $ 
in terms of eigenvectors and eigenvalues of $\hat {C}$, or use computer to 
arrange elements of matrix $\hat {X}_0 $ in vector form and solve resulting 
linear system.

The contribution to DM from a given element can be determined from Eq. (\ref{eq25})
\begin{equation}
\label{eq29}
DM_i ={\cal N}_{ii} \frac{\partial \sigma ^2(y)}{\partial {\cal N}_{ii} 
}=\vec {v}^T\frac{\partial \hat {X}_0 }{\partial \ln {\cal N}_{ii} }\vec 
{v}
\end{equation}
The topological DM contributions are
\begin{equation}
\label{eq30}
TDM_i =\vec {v}^T\frac{\partial \hat {X}_0 }{\partial {\cal N}_{ii} }\vec 
{v}
\end{equation}
Using Eqs. (\ref{eq28}) and (\ref{eq29}) one can analyze a variety of network 
connectivities. Some new effect emerging for non-tree systems are described 
next.

\subsection{Case 1: fan-out hub effect}
\label{fanout}

We now consider network shown in Figure 11A, in which all elements have the 
same variance of noise and all connections have unitary strength. Figure 11A 
shows two pathways from unit \textbf{2} to the exit unit, \textbf{6}. The 
resulting network gain from unit \textbf{2} to unit \textbf{6} is thus equal 
to two. All other units' gain at the exit is one. The contribution to DM 
from unit \textbf{2 }is thus four times larger that from other units. This 
is because noise at this unit is multiplied by a factor of two, and the 
variance of noise, by a factor of four. We conclude that there may be some 
special elements in network, which occupy hub-like positions, gaining large 
influence due to abundance of their outputs. It should be noted that fan-in 
hubs are not special from the point of view of DM in any way.

\begin{figure}[htbp]
\centerline{\includegraphics[width=4.0in]{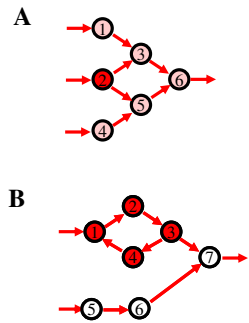}}
%\centerline{\includegraphics[width=1.6in]{dma14.eps}}
\caption{
Two cases, in which the identities of decision makers 
can be found using discrete-time approach. Variance of noise on all elements 
is the same; all network links have unitary strength. The degree of 
decision making is shown by the intensity of red. \textbf{A}, The fan-out 
effect. \textbf{B}, the temporal integrator.
}
\label{fig11}
\end{figure}

\subsection{Case 2: temporal integrator}
\label{integrator}

Let us now examine the network with a loop. Figure 11B shows such an example 
with unitary link strength and uniform noise variance, as in previous case. 
The presence of loop affects DM drastically: our discrete-time model marks 
units belonging to the loop as decision makers\footnote{ Rigorously 
speaking, the set of equations (\ref{eq28}) and (\ref{eq29}) does not have a valid 
solution for the loop with all connection equal to unity. One needs to set 
one of the connection as a parameter, $\alpha <1$, solve the equations, and 
consider the limit $\alpha \to 1 .$ }. This is easy to understand, since 
noise, generated by each unit on each time-step, cannot leave the loop and, 
therefore, builds up there without limits. Therefore the variance of noise 
in the output of element number three grows proportionally to time 
$\overline {\left[ {x_3 (t)-\overline {x_3 } (t)} \right]^2} =\overline 
{\eta ^2} t\to \infty $. Here averaging is assumed over instantiations of 
noise (trials). Thus, loop becomes the crucial decision maker. This case is 
somewhat analogous to our previous `noisy' neuron example.

What is the possible role of loops in biological networks? Why would one 
introduce such unreliable components? Loops, similar to shown in Figure 11B, 
have many useful properties. For instance, they can act as parametric memory 
systems. Indeed, imagine that responses of all units in the loop have the 
same values, equal to $x$. This could be accomplished by manipulating the 
sensory inputs. Assume that no more inputs are received from the outside of 
the system. It follows that, in the absence of noise on each element, this 
value of response will reverberate around the loop forever. This is because 
all links have unitary strength. Loops can thus memorize a graded value, 
such as $x$, functioning as parametric memory elements. 

Suppose, in addition, that a non-zero input $s$ is applied to element number 
\textbf{1} at all times. Since this element acts as a summator, its response 
on the next step is $x_1 (1)=x+s$. The signal $s$ propagates around the 
loop, and in four steps it reaches the first element again, at which time 
its response is $x_1 (5)=x+2s$. In four more steps $x_1 (9)=x+3s$. Thus, not 
only noise, but also signal can build up in the system. Therefore, a loop 
can operate as a temporal integrator. The integration is not perfect if one 
of the links has a non-unitary strength, in which case integrator becomes 
leaky (Robinson, 1989). 

Temporal integrators play special role in DM, since they act as accumulators 
of sensory information, which puts them into special position with respect 
to other areas (Gold and Shadlen, 2002). As an example, 
such is area LIP in primate visual cortex, which is involved in DM in 
direction-discrimination task 
(Shadlen and Newsome, 2001; Roitman and Shadlen, 2002; Mazurek et al., 2003). 

\subsection{Continuous-time model}
\label{continuous}

We finally consider a model, in which time runs continuously. This model has 
potential relevance to real-life networks. The responses of units satisfy 
the following equation
\begin{equation}
\label{eq31}
\frac{d\vec {x}(t)}{dt}=-\hat {A}\vec {x}(t)+\vec {\eta }(t)+\vec {s}(t).
\end{equation}
The network connectivity matrix $\hat {A}$ can be related to connection 
matrix from the discrete-time model in (\ref{eq19}) through $\hat {C}=e^{-\hat 
{A}\tau }$ (see Appendix B). Noise is defined by its cross-correlation
\begin{equation}
\label{eq32}
\overline {\eta _i (t_1 )\eta _j (t_2 )} ={\cal N}_{ij} \delta (t_1 -t_2 ),
\end{equation}
where 
\begin{equation}
\label{eq33}
\hat {{\cal N}}=\left( {{\begin{array}{*{20}c}
 {\overline {\eta _1^2 } } \hfill & \hfill & \hfill & 0 \hfill \\
 \hfill & {\overline {\eta _2^2 } } \hfill & \hfill & \hfill \\
 \hfill & \hfill & {...} \hfill & \hfill \\
 0 \hfill & \hfill & \hfill & {\overline {\eta _N^2 } } \hfill \\
\end{array} }} \right)
\end{equation}
is a diagonal cross-correlogram of noise. Eqs. (\ref{eq31})-(\ref{eq33}) are analogous 
to the discrete-time case (\ref{eq19})-(\ref{eq21}). Similarly, we define the output 
scalar and the decision variable
\[
y=\vec {v}^T \cdot \vec {x}(t)
\]
\begin{equation}
\label{eq34}
d=H(y).
\end{equation}
Here $t$ is time when the system makes the decision.

Our model is thus defined by Eqs. (\ref{eq31})-(\ref{eq34}). We will now use definition 
(\ref{eq29}) to find decision makers. As in discrete-time case we need to know the 
variance of the output variable, $\sigma ^2(y)$, after which (\ref{eq29}) leads to 
\begin{equation}
\label{eq35}
DM_i =\overline {\eta _i^2 } \frac{\partial \sigma ^2(y)}{\partial \overline 
{\eta _i^2 } }
\end{equation}
Important for us is the time-dependent correlator
\begin{equation}
\label{eq36}
\hat {X}(t_1 ,t_2 )=\overline {\vec {x}(t_1 )\vec {x}^T(t_2 )} ,
\end{equation}
which we now evaluate. Solution of (\ref{eq31}) is obtained using matrix 
exponentials 
\begin{equation}
\label{eq37}
\vec {x}(t)=\int\limits_{-\infty }^t {dt'e^{\hat {A}(t'-t)}\left[ {\vec 
{\eta }(t)+\vec {s}(t)} \right]} 
\end{equation}
If external stimulus is zero or a constant in time, due to (\ref{eq5}), the 
correlator at $t_1 >t_2 $
\begin{equation}
\label{eq38}
\hat {X}(t_1 ,t_2 )=\int\limits_{-\infty }^{t_2 } {dt'e^{\hat {A}(t'-t_1 
)}\hat {{\cal N}}e^{\hat {A}^T(t'-t_2 )}} 
\end{equation}
We seek $\hat {X}(t_1 ,t_2 )$ in the form
\begin{equation}
\label{eq39}
\hat {X}=e^{\hat {A}(t_2 -t_1 )}\hat {X}_0 ,
\end{equation}
where $\hat {X}_0 $ is equal-time cross-correlation. To find equation for 
$\hat {X}_0 $ we differentiate (\ref{eq38}) as follows
\begin{equation}
\label{eq40}
\begin{array}{l}
 \frac{\partial \hat {X}}{\partial t_2 }=\hat {A}e^{\hat {A}(t_2 -t_1 )}\hat 
{X}_0 = \\ 
 =e^{\hat {A}(t_2 -t_1 )}\hat {{\cal N}}-\hat {X}\hat {A}^T \\ 
 \end{array}
\end{equation}
We arrive thus to the following equation for $\hat {X}_0 $
\begin{equation}
\label{eq41}
\hat {A}\hat {X}_0 +\hat {X}_0 \hat {A}^T=\hat {{\cal N}}
\end{equation}
This equation is the central tool for the continuous-time theory. The 
contributions to DM from each unit are found by differentiating $\sigma 
^2(y)=\vec {v}^T\hat {X}_0 \vec {v}$ with respect to noise, as in Eq. (\ref{eq29})
\begin{equation}
\label{eq42}
DM_i ={\cal N}_{ii} \vec {v}^T\frac{\partial \hat {X}_0 }{\partial {\cal 
N}_{ii} }\vec {v}
\end{equation}
Once the same-time correlation matrix $\hat {X}_0 $ is found from Eq. 
(\ref{eq41}), cross-correlation for arbitrary time is 
\begin{equation}
\label{eq43}
\hat {X}(t_1 ,t_2 )=\left\{ {{\begin{array}{*{20}c}
 {e^{\hat {A}(t_2 -t_1 )}\hat {X}_0 ,} \hfill & {t_1 \ge t_2 } \hfill \\
 {\hat {X}_0 e^{\hat {A}^T(t_1 -t_2 )},} \hfill & {t_1 <t_2 } \hfill \\
\end{array} }} \right.
\end{equation}
This equation suggests a helpful strategy for determining noise matrix $\hat 
{{\cal N}}$. Indeed, (\ref{eq41}) and (\ref{eq43}) imply that
\begin{equation}
\label{eq44}
\hat {{\cal N}}=\left. {\frac{\partial \hat {X}(t_1 ,t_2 )}{\partial t_1 }} 
\right|_{t_1 =t_2 -\varepsilon } \left. {-\frac{\partial \hat {X}(t_1 ,t_2 
)}{\partial t_1 }} \right|_{t_1 =t_2 +\varepsilon } 
\end{equation}
Here $\varepsilon $ is infinitesimally small positive number. In other 
words, noise matrix is equal to discontinuity in time-derivative of 
cross-correlation at $t_1 =t_2 $. Since noise correlation matrix is 
diagonal, the non-zero elements are 
\begin{equation}
\label{eq45}
\overline {\eta _i^2 } =\left. {\frac{\partial X_{ii} (t_1 ,t_2 )}{\partial 
t_1 }} \right|_{t_1 =t_2 -\varepsilon } \left. {-\frac{\partial X_{ii} (t_1 
,t_2 )}{\partial t_1 }} \right|_{t_1 =t_2 +\varepsilon } 
\end{equation}
Two comments are in order here. First, noise term $\vec {\eta }(t)$ plays 
the role of input noise in (\ref{eq31}). It cannot be measured directly. Equation 
(\ref{eq44}) provides a way to single it out. Second, (\ref{eq44}) does not apply to the 
discrete-time model. Indeed, in the latter we either have $t_1 =t_2 $, or 
$t_1 =t_2 \pm 1$, etc., i.e. the condition $t_1 =t_2 \pm \varepsilon $ with 
$\varepsilon $ infinitesimally small is hard to enforce. It may happen that 
$\varepsilon \approx 1$ is acceptable due to presence of slow components in 
the circuit, such as temporal integrators. However, in general case (\ref{eq44}) 
cannot be applied to the discrete-time case. For instance, it fails 
dramatically for the case of `nematode' chain considered above.

Equations (\ref{eq41}), (\ref{eq42}), and (\ref{eq44}) represent a useful set of tools to find 
DM components for various connectivities. We present here two possible 
cases, in which decision makers can be found. They differ in what is known 
about the system.

\underline {\textbf{Scenario 1:}} Assume we know the network connectivity 
$\hat {A}$, output metrics vector $\vec {v}$, and autocorrelation for each 
unit $X_{ii} (t_1 ,t_2 )$. The steps below allow finding the 
decision makers.

\begin{enumerate}
\item Since noise matrix is diagonal, as per (\ref{eq33}), it can be found from autocorrelation using (\ref{eq44}). 
\item Solving (\ref{eq41}) allows determining $\partial \hat {X}_0 /\partial {\cal N}{ }_{ii}$, the derivative of equal-time crosscorrelation with respect to noise in each element. 
\item Decision makers are found from (\ref{eq42}).
\item Normalize contributions to DM so that $\sum\limits_i {DM_i =1} $.
\end{enumerate}

Scenario 1 does not require simultaneous measurements from all units. It 
requires the knowledge of the network connectivity however. The next 
scenario is complimentary in this respect.

\underline {\textbf{Scenario 2:}} Suppose we have measured the full 
crosscorrelation matrix $\hat {X}(t_1 ,t_2 )$ by simultaneous recordings 
from all units. Suppose also that we know how the output of the system is 
evaluated (vector $\vec {v})$. These are the steps to determine DM units.

\begin{enumerate}
\item Use (\ref{eq44}) to find noise matrix $\hat {{\cal N}}$.
\item Use (\ref{eq41}) to find the connection matrix $\hat {A}$.
\item Solve (\ref{eq41}) to calculate $\partial \hat {X}_0 /\partial {\cal N}{ }_{ii}$ for each element.
\item Use (\ref{eq42}) to find decision makers. 
\item Normalize contributions to DM so that $\sum\limits_i {DM_i =1} $.
\end{enumerate}

Both scenarios use extensive knowledge about the system, which renders them 
useless in experimental conditions. In the next subsection we discuss a way 
to bypass these limitations.

Finally, we would like to provide solution to (\ref{eq41}) using eigenbasis of 
matrix $\hat {A}$. Since $\hat {A}$ is not necessarily symmetric, a 
distinction should be made between right and left eigenvectors. The latter 
turn out to be useful for our purposes. They are defined by 
\begin{equation}
\label{eq46}
\vec {\xi }_\alpha ^+ \hat {A}=\lambda _\alpha \vec {\xi }_\alpha ^+ .
\end{equation}
Here and below Greek indexes denote numbers of eigenvalues, while Latin ones 
label spatial components of vectors and matrices. Solution of (\ref{eq41}) is

\begin{equation}
\label{eq46_5}
%X_{0ij} =\sum\limits_{{\begin{array}{*{20}c}
% {\alpha \beta \gamma \delta } \hfill \\
% {mn} \hfill \\
%\end{array} }} {\frac{\xi _{i\alpha } \xi _{j\beta }^\ast \xi _{m\gamma 
%}^\ast \xi _{n\delta } }{\lambda _\gamma +\lambda _\delta ^\ast }} \left( 
%{G^{-1}} \right)_{\alpha \gamma } \left( {G^{-1}} \right)_{\beta \delta 
%}^\ast {\cal N}_{mn} ,
X_{0ij} =\sum\limits_{\alpha \beta \gamma \delta \\ \\mn} {\frac{\xi _{i\alpha } \xi _{j\beta }^\ast \xi _{m\gamma 
}^\ast \xi _{n\delta } }{\lambda _\gamma +\lambda _\delta ^\ast }} \left( 
{G^{-1}} \right)_{\alpha \gamma } \left( {G^{-1}} \right)_{\beta \delta 
}^\ast {\cal N}_{mn} ,
\end{equation}
where
\begin{equation}
\label{eq47}
G_{\alpha \beta } =\sum\limits_i {\xi _{i\alpha }^\ast \xi _{i\beta } } 
\end{equation}
is the Gram matrix of eigenvectors. Eq. (\ref{eq46_5}) is valid if the eigenvectors form
a complete basis in the $N$-dimensional space. As follows from (\ref{eq46_5}), eigenvalues of 
$\hat {A}$ with small real part contribute to DM in a large degree. This 
justifies the use of principal component analysis when such eigenvalues are 
present. An example of such principal component is the temporal integrator 
loop in Figure 11B, which has vanishing $\lambda $. 

In case if matrix $\hat {A}$ is symmetric, its eigenvalues are real and 
eigenvectors are orthogonal. This leads to a unit Gram matrix. Then, Eq. 
(3.29) becomes more compact
\begin{equation}
\label{eq48}
X_{0ij} =\sum\limits_{
 {\alpha \beta } 
 {mn} } {\frac{\xi _{i\alpha } \xi _{j\beta }^\ast \xi _{m\alpha 
}^\ast \xi _{n\beta } }{\lambda _\alpha +\lambda _\beta }} {\cal N}_{mn} 
\end{equation}
Similar equations, called Kubo formulas, are obtained for various 
correlators in case of diffusion of particles in random media 
(Efetov, 1997). The distinguishing feature of (\ref{eq48}) is that 
a product of four eigenvectors enters the expression. Thus, propagation of 
noise in this case can be accompanied by interference between different 
pathways. An example of destructive interference of this kind is given 
below, in section \ref{stimulation}. 

Eq. (\ref{eq48}) can be further simplified. Indeed, our model uses diagonal noise 
matrices, i.e. $n=m$ in (\ref{eq48}). Suppose also that the output from the 
network occurs through one exit element number $i$, which is specified by 
taking $\vec {v}=\hat {e}_i $. In this case the use of Eq. (\ref{eq42}) gives
\begin{equation}
\label{eq49}
DM_i ={\cal N}_{nn} \sum\limits_{\alpha \beta } {\frac{\xi _{i\alpha } \xi 
_{i\beta }^\ast \xi _{n\alpha }^\ast \xi _{n\beta } }{\lambda _\alpha 
+\lambda _\beta }} .
\end{equation}
From this equation we conclude that for element $n$ to contribute to DM, an 
eigenvector should exist, which is non-zero on both unit number $n$ and exit 
unit $i$. Thus, we conclude that eigenvectors of $\hat {A}$ should be 
delocalized for broader impact of elements on the decision. This is not 
surprising in view of the mentioned analogy with the diffusion problem. In 
case if matrix $\hat {A}$ is not symmetric, the Gram matrix may be non-diagonal 
and (\ref{eq49}) cannot be used. However, the off-diagonal elements of $\hat {G}$ 
are usually smaller than diagonal ones, due to uncorrelated sign changes, 
when (\ref{eq47}) is computed with $\alpha \ne \beta $. Therefore, (\ref{eq49}) may 
apply approximately. 

\section{Analysis using stimulation}
\label{stimulation}

Stimulations with electric current add a new degree of freedom to DMA, thus 
leading to more effective ways of finding decision makers. There are two 
great advantages of the stimulation method. First, it only involves 
stimulation of a single neuron, therefore no simultaneous multiple-electrode 
measurements are required. Second, the knowledge of network connectivity is 
not needed to solve the problem. In this section we study our simple 
networks and find what stimulation strategies are consistent with our 
earlier definitions, such as Eq. (\ref{eq14}). 

We will use continuous model for concreteness (section \ref{continuous}). Consider the 
output variable $y$. It is a linear function of the inputs. It is also a 
function, which contains noise components, variable from trial to trial. The 
noise components were acquired from all units in different degree. Since 
noise in each unit is gaussian, the output variable is described by gaussian 
distribution too
\begin{equation}
\label{eq50}
\rho (y)=\frac{1}{\sqrt {2\pi \sigma ^2(y)} }
e^{-(y-\bar {y}(s))^2/2\sigma ^2(y)}
\end{equation}
In each trial a random value of $y$ is obtained, according to distribution 
(\ref{eq50}). The response of the system is equal to 1 if $y$ is positive, and 
0 otherwise. The probability to obtain response equal to 1 to given 
stimulus $s$ is given by the error function (Abramowitz and 
Stegun, 1972)
\begin{equation}
\label{eq50_5}
p_1 (s)=\int\limits_0^\infty {\rho (y)dy} =\frac{1}{2}\left[ 
{1+\mbox{erf}\left( {\frac{\bar {y}(s)}{\sigma (y)\sqrt 2 }} \right)} 
\right],
\end{equation}
whereas the probability of zero response is 
\begin{equation}
\label{eq51}
p_0 (s)=\int\limits_{-\infty }^0 {\rho (y)dy} =\frac{1}{2}\left[ 
{1-\mbox{erf}\left( {\frac{\bar {y}(s)}{\sigma (y)\sqrt 2 }} \right)} \right].
\end{equation}
Both probabilities depend upon the mean response to stimulus $\bar {y}(s)$ 
and the standard deviation $\sigma (y)$. Therefore the electric stimulation 
strategies may be based on affecting either the former or the latter. We now 
consider both of these strategies and show that affecting the mean response 
may provide misleading results, while changing the variance of response 
allows estimating contributions to DM consistently with our previous 
definitions. Thus, strategies of stimulation based on standard deviation of 
the output variable are \textit{always} correct in our simple model, independently on the 
topology of the network. This may seem a trivial consequence of definition 
(\ref{eq14}), but we will discuss it here for the sake of comparison of two 
strategies and optimizing them.

We start with the strategies of stimulation, which affect the mean response 
$\bar {y}(s)$. In our simple model this may be accomplished by injecting a 
tonic input current into a unit number $i$. Mathematically it is 
accomplished by adding extra stimulus $s_i $ to this unit in Eq. (\ref{eq31}). 
Note that in biological systems the stimulating current is alternating with 
constant amplitude (Salzman et al., 1992). The 
mean response is shifted by the stimulation, i.e.
\begin{equation}
\label{eq52}
\Delta \bar {y}=\frac{\partial \bar {y}}{\partial s_i }s_i ,
\end{equation}
where $s_i $ is the magnitude of injected tonic current. This leads to 
observable changes in the probability $p_1 $
\begin{equation}
\label{eq53}
\Delta p_1 (i)=\frac{\partial p}{\partial \bar {y}}\frac{\partial \bar 
{y}}{\partial s_i }s_i .
\end{equation}
Here $\Delta p_1 (i)$ is the change in probability of correct responses 
after unit number $i$ is electrically stimulated. Can $\Delta p_1 (i)$ be a 
measure of DM? 

We notice that $\Delta p_1 (i)$ can be either positive or negative. This 
depends on the sign of derivative $\partial \bar {y}/\partial s_i $, which 
is positive for excitatory pathway from unit $i$ to the output and negative 
for inhibitory pathway. Since contribution to DM ought to be positive, we 
cannot assume simply that $DM_i \sim \Delta p_1 (i)$. The correct 
expression, which we provide here without derivation is 
\begin{equation}
\label{eq54}
DM_i \sim \overline {\eta _i^2 } \left[ {\Delta p_1 (i)} \right]^2.
\end{equation}
This equation is understood in proportional sense, since $DM_i $ should be 
normalized to ensure that $\sum\limits_i {DM_i } =1$. Our 
investigations show that this expression is accurate for trees and is 
consistent with both our earlier definitions (\ref{eq9}) or (\ref{eq14}). Remarkably, it 
employs quantities, which can be measured in a single-electrode experiment. 
Indeed, the amplitude of noise $\overline {\eta _i^2 } $ can be found from 
autocorrelation of unit's response, using (\ref{eq44}); and $\Delta p_1 (i)$ is 
determined from behavioral changes in response to single-unit stimulation. 
This equation thus provides an approach potentially useful in practice. Does 
this relationship work for networks of arbitrary connectivity? 

Figure 12B shows a counterexample, in which a unit is stimulated, which 
results in \textit{no} change in probability of correct response (we consider trials in which 1 is the correct response throughout this section). 
This is because there are two pathways, leading from this unit to the exit, one positive and one 
negative. They have equal strength, and, therefore, compensate each other. 
On the other hand, unit number one \textit{does} participate is DM, because if a 
non-stationary stimulation/stimulus is applied, its effect on the decision 
is not zero. Thus, (\ref{eq54}) and tonic stimulation method cannot be applied to 
arbitrary circuits, such as shown in Figure 12B, to accurately reveal 
decision makers.

\begin{figure}[htbp]
\centerline{\includegraphics[width=3.0in]{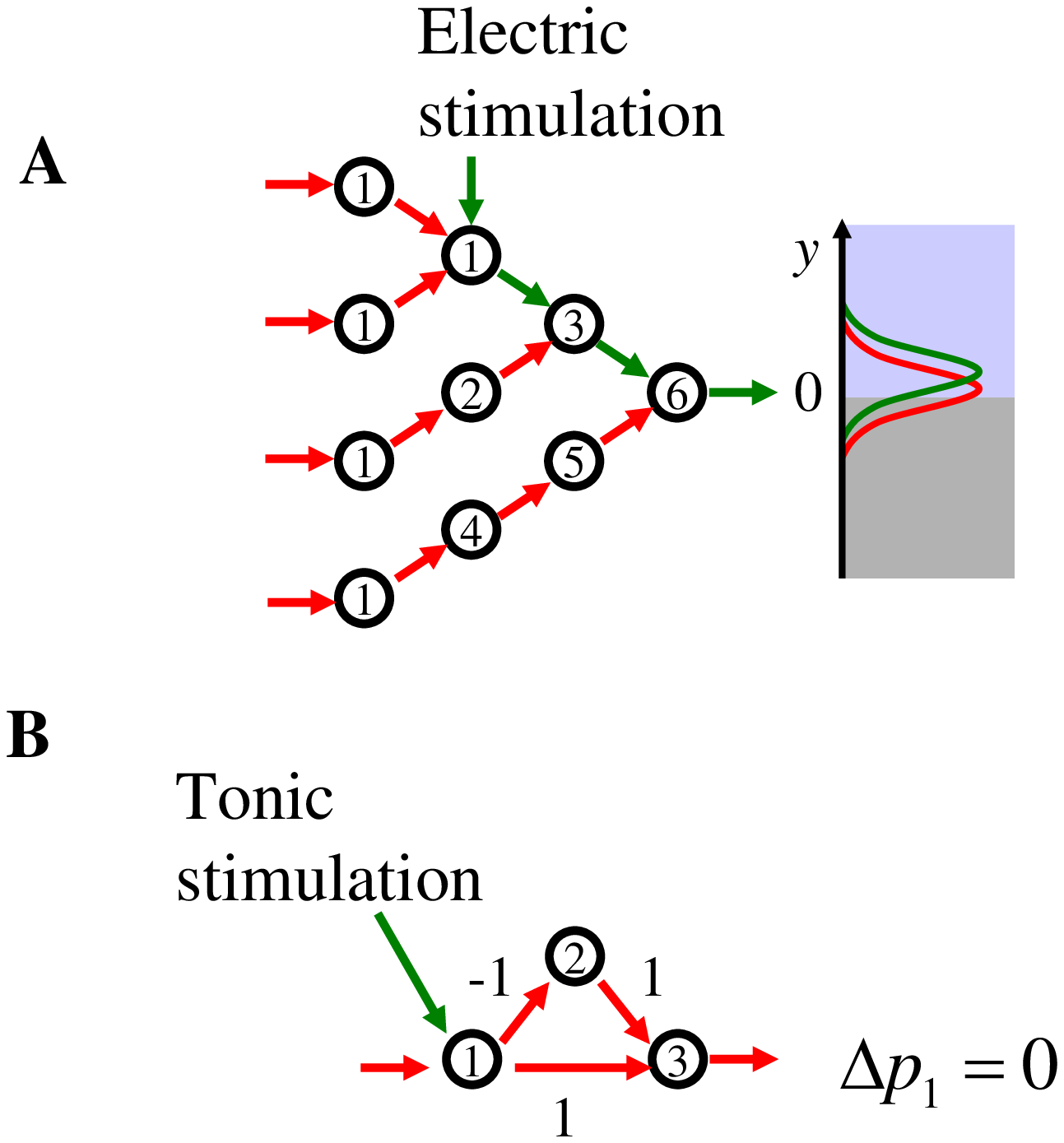}}
\centerline{\includegraphics[width=3.0in]{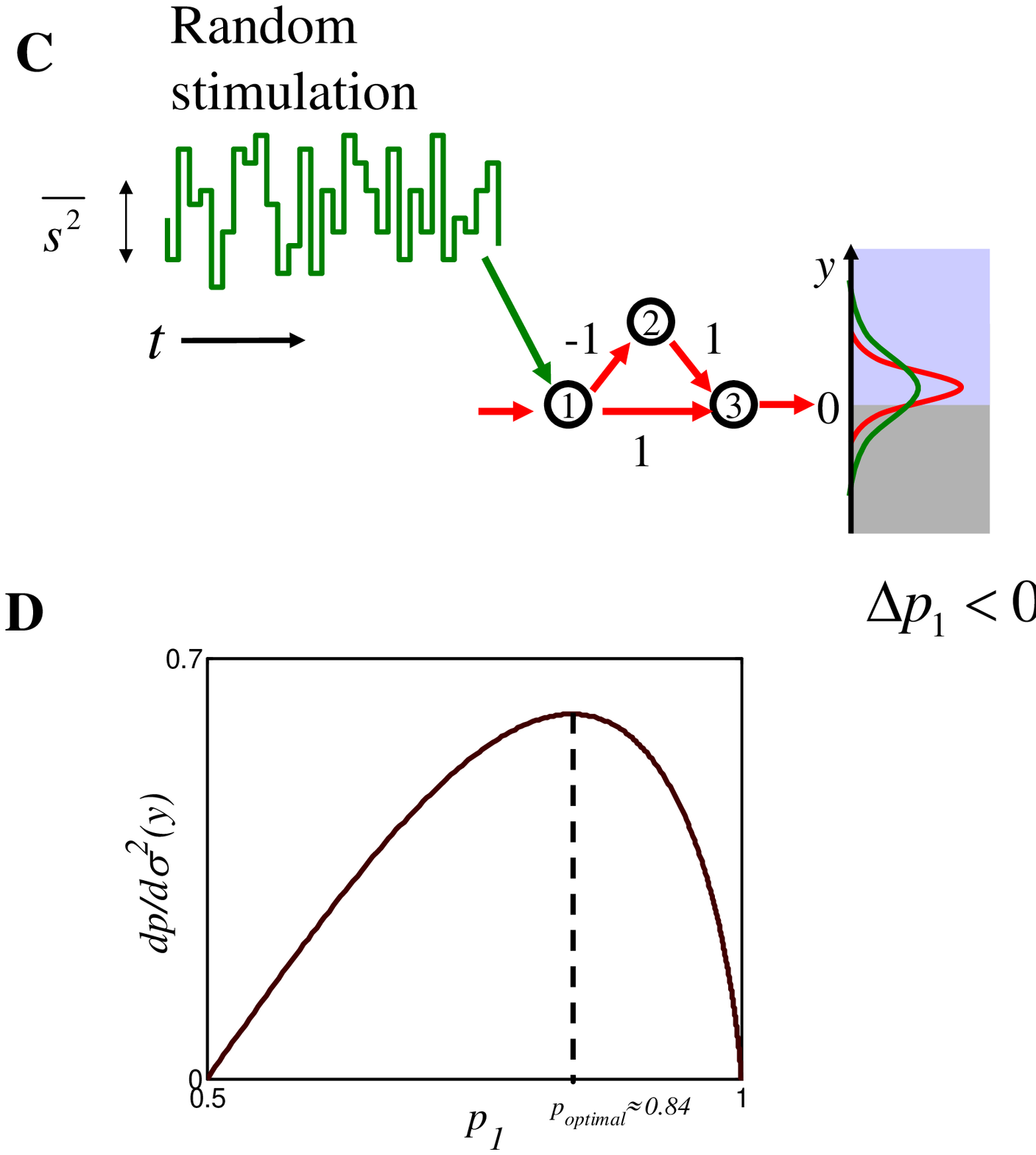}}
%\centerline{\includegraphics[width=2.67in]{dma17.eps}}
%\centerline{\includegraphics[width=1.76in]{dma18.eps}}
\caption{Finding decision makers using electric stimulation. 
\textbf{A} and \textbf{B}, tonic stimulation; \textbf{C}, random 
stimulation. \textbf{A}, tonic stimulation for trees results in shift in 
probability, which leads to correct estimation of decision making units. 
\textbf{B}, example of a circuit for which tonic stimulation leads to 
incorrect estimation of decision making, since it does not lead to the shift 
in probability. \textbf{C}, stimulation with a random current leads to 
correct estimate of decision making for networks with \textit{any} connectivity. 
\textbf{D}, for optimal performance in random stimulation paradigm, the task 
should be set so that the probability of correct responses is close to 
$p_{optimal} \approx 0.84$.
}
\label{fig12}
\end{figure}

Is there a stimulation method for finding decision making components in 
arbitrary networks? The method follows directly from the definition (\ref{eq14}) 
[or (\ref{eq35}), which is equivalent]. Indeed, when stimulating current is a 
temporal white noise, the output variable $y$ acquires a larger variance 
(Figure 12C). Hence, the derivative of output variance, entering (\ref{eq35}) can 
be calculated operationally, by injecting a distracter current. More 
precisely, if the variance of stimulating current applied to unit $i$ is 
$\overline {s_i^2 } $ the derivative entering definition (\ref{eq35}) is 
\begin{equation}
\label{eq55}
\frac{\partial \sigma ^2(y)} {\partial \overline {\eta _i^2 } } = 
\frac{\Delta \sigma^2 (y)} {\overline {s_i^2 } }.
\end{equation}
In practice one has no access to the variable $y$, so one cannot measure 
directly the change in variance $\Delta \sigma^2 (y)$. Instead, one could 
measure the change in the probability of correct responses under the 
influence of distracting current. Indeed, from (\ref{eq50_5}) we obtain
\begin{equation}
\label{eq56}
\Delta p_1 (i) = \frac{\partial p_1 } {\partial \sigma^2 (y) } \Delta \sigma^2 (y)
\end{equation}
Combining the last two equations we obtain for the important derivative
\begin{equation}
\label{eq57}
\frac{\partial \sigma ^2(y)}{\partial \overline {\eta _i^2 } }=\frac{\Delta 
p_1 (i)}{\overline {s_i^2 } }\left( {\frac{\partial p_1 }{\partial \sigma 
^2(y)}} \right)^{-1}
\end{equation}
Since the probability of correct responses always decreases under the 
influence of distracters, the derivative $\partial p_1 /\partial \sigma 
^2(y)$ is a negative constant. It is the same for all units. We arrive 
therefore to the expression for contributions to DM, which follows from 
(\ref{eq35})
\begin{equation}
\label{eq58}
DM_i \sim -\overline {\eta _i^2 } \frac{\Delta p_1 (i)}{\overline {s_i^2 } 
}
\end{equation}
Here $\Delta p_1 (i)$ is the decrease in probability of correct responses 
produced by electric stimulation with variance of the random current equal 
to $\overline {s_i^2 } $. The variance of noise on each unit $\overline {\eta 
_i^2 } $ can be found from autocorrelation using (\ref{eq45}). This procedure 
works for any topology in our simplified model. It should be noted here that 
if noise is not entirely white or cannot be considered white, (\ref{eq45}) cannot 
be used directly and should be replaced by an expression reflecting the 
spectral characteristics of noise appropriate for the system under 
investigation. Thus, if noise is provided by other parts of the network, its 
dynamic features may be more complex. Therefore, (\ref{eq45}) may not apply 
directly to the `hidden pathway' example given in the end of section \ref{conclusions}.

The procedure, which we just described, permits further optimization. 
Indeed, imagine that the probability of correct responses is exactly 
$\raise.5ex\hbox{$\scriptstyle 1$}\kern-.1em/ 
\kern-.15em\lower.25ex\hbox{$\scriptstyle 2$} $. Adding distracting 
stimulation current will not change this probability, i.e. $\Delta p_1 
(i)=0$ no matter what unit is stimulated. In the opposite limiting case when 
$p_1 \approx 1$, the effect of distracter on performance is exponentially 
small. Hence, behavioral response to stimulation has an optimum between $p_1 
=1/2$ and $1$. To find the optimum we observe from (\ref{eq56}) that $\Delta p_1 $ 
is maximum for the same variation in $\Delta \sigma ^2(y)$ when $\partial 
p_1 /\partial \sigma ^2(y)$ is maximum. We therefore plot the latter 
derivative as a function of $p_1 $ in Figure 12D. We indeed observe a 
maximum at the value of probability of correct responses close to
\begin{equation}
\label{eq59}
p_{optimal} \approx 0.841
\end{equation}
To summarize, the following scenario describes algorithm for finding 
contributions to DM using random stimulation.

\underline {\textbf{Scenario 3:}} Assume that we \textit{do not know} the network connectivity 
$\hat {A}$ and output metrics vector $\vec {v}$; but we know autocorrelation 
for each unit $X_{ii} (t_1 ,t_2 )$. The steps below allow finding the 
decision makers.

\begin{enumerate}
\item Prepare stimulus so that the probability of correct responses is close to the value given by (\ref{eq59}).
\item Stimulate one unit with random current, whose variance is $\overline {s_i^2 } $, and measure the decrease in probability of correct responses $\Delta p_1 $.
\item Record autocorrelation and evaluate noise variance $\overline {\eta _i^2 } $ for this unit using (\ref{eq45}). 
\item Find contribution to DM for this unit using equation (\ref{eq58}).
\item Repeat steps 1 through 4 for all units in the system.
\item Normalize contributions to DM so that $\sum\limits_i {DM_i =1} $.
\end{enumerate}

\section{Discussion}
\label{discussion}

In this work we defined decision makers in networks, which behave in a well-defined fashion.
As with any definition, there is certain degree of arbitrariness in our study, 
since this is the first mathematical study of this sort.
We had to make choices about the features of decision making we 
were attempting to describe as well as about the way they were quantified. 
We demonstrated these features in a set of examples.
Future studies will show if these features can be used as a basis of a more complete 
model-independent theory.

In this study we postulated that variability and noise, causally linked to decisions, 
are the chief descriptors of DM. Although this point may seem paradoxical we 
suggest three arguments in its favor. First, variability may reflect 
additional information needed to make a decision in case of uncertainty. 
Such may be inputs from other modalities, memories, or some other relevant 
modulatory inputs, supplying e.g. emotional condition of the subject 
or changing utility values (Figure 8). Second, many behaviors, such as C-start 
escape responses in fish (Eaton and Emberley, 1991) and 
other organisms (Glimcher, 2003), have stochastic character. 
This makes the task of pursuer more difficult.
Such unpredictable behaviors are reproduced in our model if the sensory input is weak 
or in the small signal-to-noise ratio case. Third, 
the goal of DM is to dissipate sensory information, as suggested in the 
introduction (Figure 1A), whereby an analog multi-dimensional stimulus space 
is reduced to a discrete space of several decisions. We argue that 
this transformation is facilitated by noise.

We have studied the problem of finding decision making units in 
networks of various connectivities. This path took us from simple linear chains, for which 
the information-theoretical (IT) approach was found to be effective, to trees, and, finally to   
an alternative definition of decision makers, based on propagation of noise in networks. 
This latter definition is valid in networks of arbitrary topology. 
All these approaches are equivalent,
when they can be compared, but include progressively broader classes of networks.
As a practical application for the alternative definition we 
considered the problem of electric stimulation in the surrogate networks and 
showed a way of determining DM contributions for arbitrary networks using 
stimulation with random current. Our findings are summarized in Figures 13 and 14.

\begin{figure}[htbp]
\centerline{\includegraphics[width=3.2in]{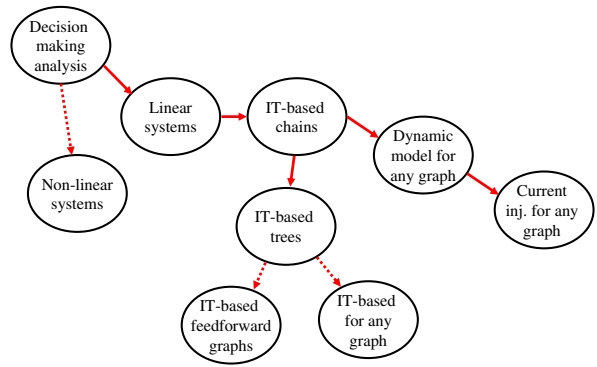}}
\caption{
The cluster of problems covered in this study. 
Solid/dashed arrows show the derivations performed here/yet to be confirmed 
or denied. IT stands for information-theoretical.
}
\label{fig13}
\end{figure}

Although we studied networks of complex connectivity,
the model describing a single network element was quite simple. 
Not all of the units are linear, of course, since DM is a non-linear task 
(Figure 1A). However, our model is essentially based on linear elements. The 
motivation for this model is that it is easy to analyze. The study of 
simple models is a necessary step before analysis proceeds any further. 
Once the methodological issues are resolved for simpler models, complex 
non-linear systems can be studied in the same paradigm. 
One of important questions resolved here is that a completely linear element 
can be a decision maker, despite the presence of non-linear units in the 
network. Thus, nonlinearity is not a necessary attribute of DM. This question 
would be impossible to answer for more realistic system, since in practice 
all units contain nonlinearities. 

%In this study we obtained identities of decision makers using correlations 
%with the response. An alternative approach would be to consider correlations 
%with sensory inputs and to surmise that the points, where such correlations 
%disappear are responsible for making decisions. It is clear, however, that 
%some pathway between the unit and motor response is necessary for the unit 
%to impact DM. If such pathway exists, our analysis can be applied. If the 
%pathway does not exist, the approach, involving sensory correlations 
%produces misleading results. On the other hand, it is possible that a 
%combined sensory-motor approach would be more useful. This question warrants 
%further investigation. 

\begin{figure}[htbp]
\centerline{\includegraphics[width=3.2in]{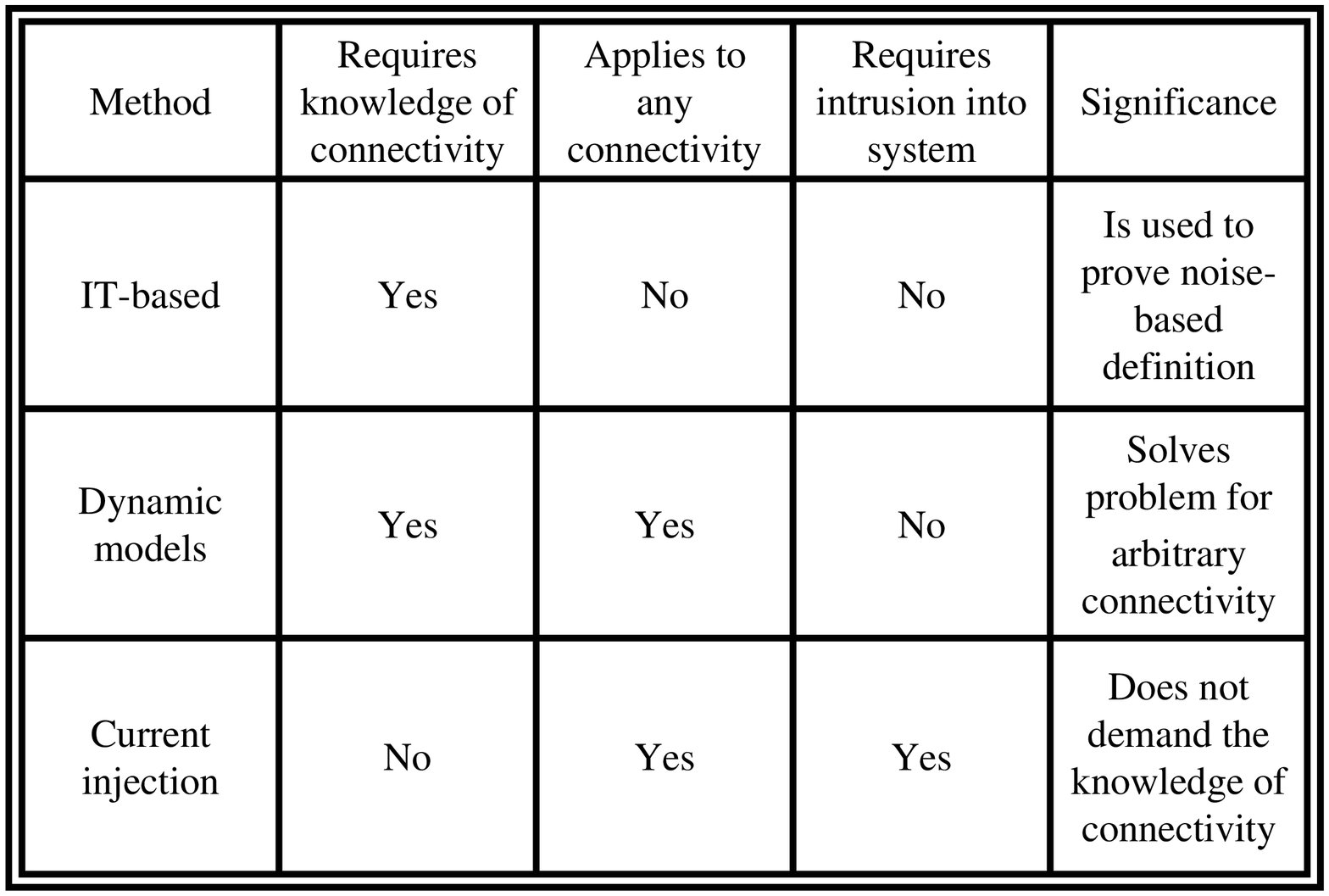}}
\caption{Comparison between different approaches studied here.}
\label{fig20}
\end{figure}

Decision making task, as formulated in Figure 1A, is similar to general 
object discrimination task. Representation of motor response in our model is 
not distinguishable mathematically from the representation of abstract 
object/decision category (Horwitz and 
Newsome, 1998; Shadlen and Newsome, 2001). The latter does not necessarily 
lead to a motor command. Thus, our analysis may uncover the identities of 
units responsible for categorization of sensory inputs. In terms of this 
analysis we emphasize the distinction between units representing the object 
category and the units in which this representation is actually formed. The 
former are analogous to motor units in the decision task, while the latter 
are similar to decision makers. As follows from this study, the analysis is 
dependent upon the topology of the network involved. For simple linear 
sensory chains our conclusion is that the \textit{first} unit, spatially or temporally, in 
which the representation of the object is correlated with final outcome of 
the discrimination process, is responsible for casting the stimulus in one 
of the abstract classes. In case of recurrent networks a more detailed 
quantitative analysis is needed to draw conclusions about identities of 
categorizing units. Thus, DMA may find a broader use in identifying units 
representing abstract object's percepts. 

A special care should be taken in distinguishing the DM task from the 
sensory discrimination task. It may occur that in the same experiment these 
tasks are performed by different populations of neurons. An example is given 
by (Salinas and Romo, 1998). They discovered a population of M1 
neurons responding differentially to two categories of tactile stimuli. Some 
of these neurons did not respond, when the same behavior was guided by 
visual cues. This observation is consistent with these neurons performing 
sensory discrimination of tactile stimuli, 
while some other population making decisions about the actual motor response.  
Our mathematical analysis is general enough to include both 
of these functions. Thus, if correlations with motor response are 
studied, it will result in the decision makers; while when the correlations 
with percepts are investigated, DMA should provide the identities of 
discriminating elements.

We suggest that DMA may be relevant to other biological systems. 
Possible applications may include the analysis of molecular networks, such 
as genetic regulatory or protein binding networks; finding decision makers 
in compartmental models of dentritic trees (Poirazi and Mel, 
2001); studies of neural networks and structural networks of connectivities 
between different brain areas; and analysis of social networks.

\section{Conclusion}
\label{gen_conclusion}

In this study we define network elements responsible for making decisions. 
We obtain two equivalent definitions. According to one, decisions are made 
by elements, in which correlations with the decision are first formed. 
According to the second definition, decision making activity is measured by 
the impact of variability in given unit on the response. We give examples of 
network motifs, especially potent from decision making prospective, such as 
fan-out hubs and recurrent loops. The latter can function as temporal 
integrators of sensory inputs. We also study how electric stimulations can 
reveal decision making components. We conclude that stimulations with 
time-varying random current produce correct results for all network 
topologies.

\appendix

\section{The linear chain model.}
\label{appendixa}

Here we solve a more general version of linear chain model than considered 
in the text. The responses of neighboring neurons are related linearly 
\begin{equation}
x_i =C_{i-1} x_{i-1} +\eta _i
\label{Eqa1}
\end{equation}
This is a generalization of (\ref{eq1}). The response of the $n$th unit is 
\begin{equation}
x_n =\sum\limits_{i=1}^n {\alpha _{ni}\eta _i } +\alpha _{n0}x_0
\label{Eqa2}
\end{equation}
where coefficients $\alpha _{ni} =C_{n-1} C_{n-2} \ldots C_i $, $\alpha 
_{nn} =1$. The external signal $x_0 $ is assumed to be zero in this 
appendix, due to (\ref{eq5}). For the last element in the chain we have 
\begin{equation}
x_N =\sum\limits_{i=1}^N {\alpha _{Ni}\eta _i } .
\label{Eqa3}
\end{equation}

Comparing (\ref{Eqa2}) and (\ref{Eqa3}) we conclude that 
\begin{equation}
x_N =\alpha _{Nn} x_n +\xi ,
\label{Eqa4}
\end{equation}
where $\xi $ is a variable, which describes noise in the networks downstream 
from unit $n$. It is, thus, uncorrelated with $x_n $. This is where 
tree-like topology enters our solution, since in case of loops, $x_n $ and 
$\xi $ are correlated. Our goal now is to calculate MI between the decision 
variable $d=H(x_N )$ and $x_n $. We will use the definition for MI 
\begin{equation}
MI(d,x_n )=\sum\limits_{d=0,1} {\int\limits_{-\infty }^\infty {dx_n \rho 
\left( {d,x_n } \right)\log _2 \left[ {\frac{\rho \left( {d,x_n } 
\right)}{\rho (d)\rho (x_n )}} \right]} }
\label{Eqa6}
\end{equation}
Here $\rho \left( d \right)=1/2$, since there is no signal;
\begin{equation}
\rho \left( {x_n } \right)=\exp \left( {-x_n^2 /2\overline {x_n^2 } } 
\right)/\left( {2\pi \overline {x_n^2 } } \right)^{1/2}
\label{Eqa8}
\end{equation}
and
\begin{equation}
\rho (d;x_n )=\frac{\rho (x_n )}{2}\left[ {1\pm \mbox{erf}\left( 
{\frac{\alpha _{Nn} x_n }{\sigma \left( \xi \right)\sqrt 2 }} \right)} 
\right].
\label{Eqa9}
\end{equation}
The upper/lower sign is assumed for $d=0$ or $1$ in (\ref{Eqa9}); $\sigma (\xi )$ 
is the standard deviation of Gaussian variable $\xi $ defined in (\ref{Eqa4}). The 
expression for MI (\ref{Eqa6}) results in 
\begin{equation}
\begin{array}{l}
 MI_n =M(s_n ) \\ 
 M(s_n )=\frac{1}{\sqrt \pi }\int\limits_{-\infty }^\infty {dze^{-z^2}\left[ 
{1+\mbox{erf}\left( {zs_n } \right)} \right]} \log _2 \left[ 
{1+\mbox{erf}\left( {zs_n } \right)} \right] \\ 
 s_n =\sigma (\alpha _{Nn} x_n )/\sigma (\xi ). \\ 
 \end{array}
\label{Eqa11}
\end{equation}
MI is therefore a function of signal-to-noise ratio $s_n$. Inversely, 
\begin{equation}
s_n^2 =\frac{\alpha _{Nn}^2 \overline {x_n^2 } }{\overline {\xi ^2} }=\left[ 
{M^{-1}(MI_n )} \right]^2
\label{Eqa12}
\end{equation}
On the other hand, (\ref{Eqa4}) leads to 
\begin{equation}
\overline {x_N^2 } =\alpha _{Nn}^2 \overline {x_n^2 } +\overline {\xi ^2}
\label{Eqa13}
\end{equation}
Solving (\ref{Eqa12}) and (\ref{Eqa13}) with respect to $\alpha _{Nn}^2 \overline {x_n^2 }$ we have 
\begin{equation}
\frac{\alpha _{Nn}^2 \overline {x_n^2 } }{\overline {x_N^2 } }=\frac{\left[ 
{M^{-1}(MI_n )} \right]^2}{1+\left[ {M^{-1}(MI_n )} \right]^2}\equiv F(MI_n 
)
\label{Eqa14}
\end{equation}
Function $M^{-1}$ here is inverse to $M$ defined in (\ref{Eqa11}). Function $F(MI)$ 
numerically calculated from (\ref{Eqa11}) and (\ref{Eqa14}) is shown in Figure 5. Lastly, 
we recall that variances $\alpha _{Nn}^2 \overline {x_n^2 } $ are related to 
the strength of noise $\overline {\eta _i^2 } $ through (\ref{Eqa2}). We have 
\begin{equation}
\displaystyle
\alpha _{Nn}^2 \overline {x_n^2 } =\sum\limits_{i=1}^n {\alpha _{Ni}^2 
\overline {\eta _i^2 } }
\label{Eqa15}
\end{equation}
Eqs. (\ref{Eqa14}) and (\ref{Eqa15}) are used below to prove a variety of statements about 
function $F(MI)$ used in the main text.

\subsection{In the uniform noise example $F(MI)$ is a linear function of position in the chain.} 

In this case $C_1 =\ldots =C_{N-1} =1$, 
and, consequently, $\alpha _{N1} =\ldots =\alpha _{NN} =1$. Noise variance 
is the same on every node, i.e. $\overline {\eta _i^2 } \equiv \eta^2 $. 
As follows from (\ref{Eqa15}) $\overline {x_n^2 } =\eta ^2n$, which results in 
\begin{equation}
F(MI_n )=n/N
\label{Eqa16}
\end{equation}
It follows that contributions to DM defined by (\ref{eq9}) are the same for all 
units.

\subsection{In the `loud' neuron example the contributions of units upstream 
from the strong link are larger by a factor of $K^2$ than 
contribution from the downstream units.}

In this case $\alpha _{1...k} =K$, 
while $\alpha _{k+1...N} =1$, assuming that the link from unit $k$ to $k+1$ 
is strengthened. In the example in the text $k=5$ [cf. (\ref{eq10})]. Eq. (\ref{Eqa15}) 
leads us to the values for variances of responses
\begin{equation}
\displaystyle
\alpha _{Nn}^2 \overline {x_n^2 } =\left\{ {{\begin{array}{*{20}c}\displaystyle
 {\displaystyle \eta ^2K^2n,} \hfill & {n\le k} \hfill \\
 {\eta ^2K^2k+\eta ^2\left(\displaystyle {n-k} \right),} \hfill & {n>k} \hfill \\
\end{array} }} \right.
\label{Eqa17}
\end{equation}
Applying (\ref{Eqa14}) we obtain the expression for $F(MI)$
\begin{equation}
F(MI_n )=\left\{ {{\begin{array}{*{20}c}\displaystyle
 {\displaystyle \frac{K^2n}{N-k+K^2k},} \hfill & {n\le k} \hfill \\
 {\displaystyle \frac{n-k+K^2k}{N-k+K^2k},} \hfill & {n>k} \hfill \\
\end{array} }} \right.,
\label{Eqa18}
\end{equation}
which is a piece-wise linear function of $n$. Eq.~(\ref{eq9}) determines 
contributions to DM as
\begin{equation}
DM_n =\left\{ {{\begin{array}{*{20}c}
 {\displaystyle \frac{K^2}{N-k+K^2k},} \hfill & {n\le k} \hfill \\
 {\displaystyle \frac{1}{N-k+K^2k},} \hfill & {n>k} \hfill \\
\end{array} }} \right.
\label{Eqa19}
\end{equation}
This confirms that the upstream units ($n\le k)$ are $K^2$ times more potent 
than the downstream ones ($n>k)$. 

\subsection{Two definitions of contribution to DM using derivative of 
$F(MI)$ (\ref{eq9}) and the impact of noise (\ref{eq14}) are equivalent. }

Let 
us start by determining decision makers from definition (\ref{eq14}). According to 
(\ref{Eqa3})
\begin{equation}
\overline {x_N^2 } =\sum\limits_{i=1}^N {\alpha _{Ni}^2 \overline {\eta _i^2 
} } .
\label{Eqa20}
\end{equation}
Definition (\ref{eq14}) gives
\begin{equation}
DM_i \propto \overline {\eta _i^2 } \frac{\partial \overline {x_N^2 } 
}{\partial \overline {\eta _i^2 } }=\alpha _{Ni}^2 \overline {\eta _i^2 } .
\label{Eqa21}
\end{equation}
After normalization we obtain
\begin{equation}
DM_i =\frac{\alpha _{Ni}^2 \overline {\eta _i^2 } }{\overline {x_N^2 } }.
\label{Eqa22}
\end{equation}
Let us derive the same result from (\ref{eq9}). As follows from (\ref{Eqa14}) 
\begin{equation}
\begin{array}{l}
\displaystyle F(MI_n )-F(MI_{n-1} )=\frac{1}{\overline {x_N^2 } }\left( {\alpha _{Nn}^2 
\overline {x_n^2 } -\alpha _{N\mbox{ }n-1}^2 \overline {x_{n-1}^2 } } 
\right) \\ 
 \displaystyle =\frac{\alpha _{Nn}^2 }{\overline {x_N^2 } }\left( {\overline {x_n^2 } 
-C_{n-1}^2 \overline {x_{n-1}^2 } } \right)=\frac{\alpha _{Nn}^2 \overline 
{\eta _n^2 } }{\overline {x_N^2 } } \\ 
 \end{array}
\label{Eqa23}
\end{equation}
This proves the equivalence of (\ref{eq9}) and (\ref{eq14}), since the result is 
identical to (\ref{Eqa22}).

\section{Connection between discrete- and continuous-time models.}
\label{appendixb}

In this section we show that the discrete-time model can be derived from 
continuous-time model. Starting from equation (\ref{eq37}) for the unit responses 
in the continuous case we obtain the relation for solutions at two different 
time points separated by the time-interval $\tau$, analogous to (\ref{eq19}) in 
the discrete-time description. Then we show that in the limiting case $\tau \to 0$ two descriptions are equivalent.

From (\ref{eq37}) we obtain
\begin{equation}
\vec {x}(t+\tau )=e^{-\hat {A}\tau }\vec {x}(t)+\int\limits_t^{t+\tau } 
{e^{-\hat {A}(t+\tau -{t}')}\vec {\eta }({t}')d{t}'}.
\label{Eqb1}
\end{equation}
This equation can be rewritten as $\vec {x}(t+\tau )=\hat {C}\vec {x}(t)+{\vec {\eta }}'(t)$, 
where 
\begin{equation}
\hat {C}=e^{-\hat {A}\tau }\approx \hat {I}-\hat {A}\tau .
\label{Eqb2}
\end{equation}
Thus it has the same form as (\ref{eq19}). Using (\ref{eq32}) we obtain that the new noise cross-correlation matrix
\begin{equation}
{\hat {{\cal N}}}'=\int\limits_0^\tau {e^{-\hat {A}{t}'}\hat {{\cal 
N}}e^{-\hat {A}^T{t}'}d{t}'} .
\label{Eqb3}
\end{equation}
The solution of the continuous-time problem satisfies the equations of the 
discrete-time model for an arbitrarily large time interval $\tau $, but 
the new noise cross-correlation matrix ${\hat {{\cal N}}}'$ is non-diagonal 
in this case. In the limiting case $\tau \to 0$ it becomes 
diagonal. Indeed (\ref{Eqb3}) implies that in this limit
\begin{equation}
{\hat {{\cal N}}}'=\hat {{\cal N}}\tau \quad ,
\label{Eqb4}
\end{equation}
which is diagonal by the definition of the continuous-time model. 
Here we kept only terms linear in $\tau $. Thus, in this limit the matrix ${\hat {{\cal N}}}'$ is diagonal as needed in our 
formulation of discrete-time model. One can also derive (\ref{eq41}) from (\ref{eq28}) 
using (\ref{Eqb2}) and (\ref{Eqb4}) and taking the limit $\tau \to 0$.

\end{document}